\documentclass[3p,twocolumn]{elsarticle}

\usepackage{mathrsfs}
\usepackage{amsmath,amsthm}  
\usepackage{amssymb}
\usepackage{color}

\usepackage{ulem}

\def\I{\mathrm{i}}                  
\def\D{\mathrm{d}}                  
\newcommand{\ket}[1]{|\kern.3ex#1\kern.3ex\rangle}
\newcommand{\bra}[1]{\langle\kern.3ex #1 \kern.3ex|}
\newcommand{\mean}[1]{\left\langle #1 \right\rangle} 
\newcommand{\smean}[1]{\langle #1 \rangle} 

 \newcommand{\derivp}[2]{\frac{\partial #1}{\partial #2}}
 
 \newcommand{\EXP}[1]{{\mathrm{e}}^{#1}}         

 \newcommand{\re}{\mathop{\mathrm{Re}}\nolimits}      

\newcommand{\diagram}[3]{\raisebox{#3}{\includegraphics[scale=#2]{#1}}}

\newcommand\ab{{\alpha\beta}}

\newcommand\mn{{\mu\nu}}

\def\xiloc{\xi_\mathrm{loc}}
\def\sw{s}

\def\Gintern{\mathcal{G}}

\begin{document}

\title{Four-terminal resistances in mesoscopic networks of metallic wires~:\\
       Weak localisation and correlations \tnoteref{t1}} 
\tnotetext[t1]{Contribution to a special issue ``\textit{Frontiers in quantum electronic transport -- in memory of Markus B\"uttiker}''}

\author[LPTMS,LPS]{Christophe Texier}
\ead{christophe.texier@u-psud.fr}
\author[LPS]{Gilles Montambaux}
\ead{gilles.montambaux@u-psud.fr}
  
\address[LPTMS]{Laboratoire de Physique Th\'eorique et Mod\`eles Statistiques, Universit\'e Paris-Sud, UMR 8626 du CNRS, B\^at. 100, 91405 Orsay, France}
\address[LPS]{Laboratoire de Physique des Solides, Universit\'e Paris-Sud, UMR 8501 du CNRS, B\^at. 510, 91405 Orsay, France}

\begin{abstract}
  We consider the electronic transport in multi-terminal mesoscopic networks of weakly disordered metallic wires.
  After a brief description of the classical transport, we analyze the weak localisation (WL) correction to the four-terminal resistances, which involves an integration of the Cooperon over the wires with proper weights. We provide an interpretation of these weights in terms of classical transport properties.
  We illustrate the formalism on examples and show that weak localisation to four-terminal conductances may become large in some situations.
  In a second part, we study the correlations of four-terminal resistances and show that integration of Diffuson and Cooperon inside the network involves the same weights as the WL. The formulae are applied to multiconnected wire geometries.
\end{abstract}





\maketitle


\graphicspath{{./FigBut4term/}}


\section{Introduction}

Classical laws of transport on electrical networks have been established by Gustav Kirchhoff in 1845. They rely on three fundamental hypotheses. Two of them are energy and charge conservation leading respectively to the so-called voltage and current laws.
The third one is Ohm's law which states that the current $I_{\mu\nu}$ along a wire $(\mu\nu)$ of the network is proportional to the voltage drop between the vertices $\mu$ and $\nu$  connected by this wire~:
\begin{equation}
  \label{eq:Ohm}
  I_{\mu\nu}= \frac{\sigma_0\sw}{l_{\mu\nu}}(V_\mu - V_\nu)  
  \:, 
\end{equation}
$l_{\mu\nu}$ being the length of the wire, $\sw$ its cross section and $\sigma_0$ the Drude conductivity.
We now understand that this third hypothesis relies on the assumption of diffusive motion of the charge carriers at the microscopic level, within a classical description.
In particular, it is not appropriate to describe quantum effects (like Aharanov-Bohm effect) or non-diffusive regime (ballistic regime, quantum Hall effect). 
After the fundamental breakthrough proposed by Rolf Landauer to describe the electrical conductance as a transmission coefficient, a generalization of laws of transport beyond classical transport was highly desirable. 
It has been formalized in a beautiful work by Markus B\"uttiker~(for reviews,  see Refs.~\cite{But92,Dat95,Imr97,NazBla09}).

If we restrict ourselves to the regime of linear transport, 
a convenient description is to start by introducing the conductance matrix $G$ which relates the currents at the contacts (also called terminals) of the circuit to the values of the voltage at these terminals 
\begin{equation}
  \label{eq:IdeV}
  I_\alpha = \sum_\beta G_{\alpha\beta} V_\beta 
  \:.
\end{equation}
This relation  is completely general in the limit of linear transport.
The two first Kirchhoff's laws imply~:
\begin{equation}
  \sum_\alpha G_{\alpha\beta}=\sum_\beta G_{\alpha\beta}=0
  \:,
\end{equation}
expressing the conservation of current, and the invariance of the current distribution against a global shift of the potentials (gauge invariance).
As a generalization of Landauer's formula,  elements of the conductance matrix are related to transmission coefficients
\begin{equation}
  G_{\alpha\beta}= - {2_s e^2 \over h} T_{\alpha\beta} 
  \quad \mbox{for} \quad 
  \beta\neq \alpha 
  \:,
\end{equation}
where the factor $2_s$ stands for the spin degeneracy.
At this level, the formalism is completely general and no hypothesis is made on the nature of electrical transport which is totally encoded in the transmission coefficients $T_{\alpha\beta}$.
These coefficients are related to the scattering matrix which may be determined explicitly within specific models.
For example, in the regime of the integer quantum Hall effect, the current is carried by edge states which makes the problem effectively one-dimensional and allows for a simple construction of the scattering matrix \cite{But88,But91}.
More generally, the scattering matrix may be constructed efficiently by assuming strictly one-dimensional character, like in a network of strictly one-dimensional (1D) wires \cite{TexMon01}.
Another powerful approach applies to devices in which the electron dynamics can be considered as ergodic, leading to a random matrix formulation of quantum scattering \cite{Bee97}. 
In the present article, we consider the case of metallic samples made of weakly disordered wires such that the electron dynamics is \textit{diffusive}, as it is the case in narrow metallic wires deposited on a substrate \cite{WasWeb86} (see also the recent experiments \cite{SchMalMaiTexMonSamBau07}), or wires etched in a two-dimensional-electron gas~\cite{FerAngRowGueBouTexMonMai04,FerRowGueBouTexMon08,CapTexMonMaiWieSam13}.

\begin{figure}[!ht]
\centering
\includegraphics[width=0.35\textwidth]{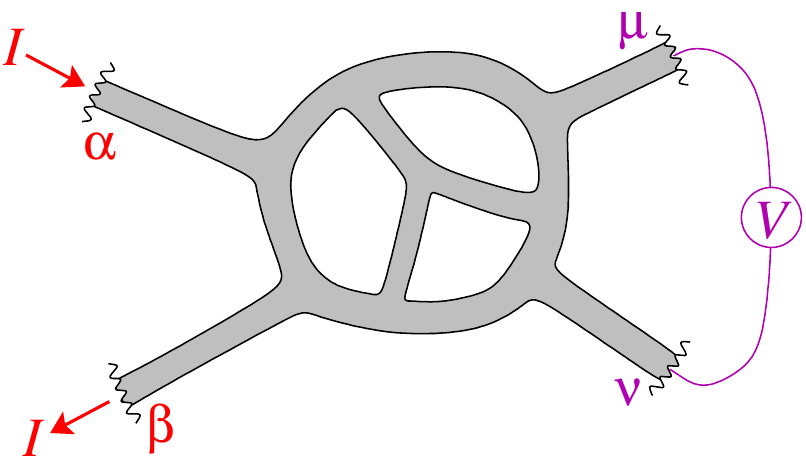}
\caption{\textit{The four-terminal resistance $\mathcal{R}_{\alpha\beta,\mu\nu}$ is the ratio of the voltage between two contacts $\mu$ and $\nu$ and the current injected at contact $\alpha$ and collected at contact $\beta$.}}
\label{fig:4TR}
\end{figure}

B\"uttiker emphasized the importance of the measurement process when a quantum circuit is connected to the outside macroscopic world~\cite{But87}. The measured resistance is not only a property of the system itself but also depends on the way it is connected to the outside world.  Moreover, although the concept of conductance is natural from a theoretical point of view,  experiments most  frequently deal with voltage measurements~: current is injected and collected at two specific contacts and voltages are measured at any pair on contacts  playing the role of {\it voltage probes} (Fig.~\ref{fig:4TR}).
Therefore, the relevant quantities characterizing the response of the device are the four-terminal resistances defined as
\begin{equation}
  \label{eq:Def4TR}
  \mathcal{R}_{\alpha\beta,\mu\nu} = \frac{V_\mu-V_\nu}{I}
  \hspace{0.25cm}
  \mbox{with }
  \begin{cases}
     I_\alpha=I\\
     I_\beta=-I\\
     I_\lambda=0\ \forall\,\lambda \neq\alpha,\,\beta
  \end{cases}
\end{equation}
By appropriate inversion of relation (\ref{eq:IdeV}), B\"uttiker could relate the four-terminal resistances to the elements of the conductance matrix, therefore to the transmission coefficients~\cite{But86a,But88a}~:
\begin{equation}
  \label{eq:Buttiker1986}
  \mathcal{R}_{\alpha\beta,\mu\nu} = \frac{h}{2_se^2}
  \frac{T_{\mu\alpha}T_{\nu\beta}-T_{\mu\beta}T_{\nu\alpha}}{\mathcal{D}}
  \:,
\end{equation}
where $\mathcal{D}$ is any minor of the dimensionless conductance matrix.
The expression is valid when all indices are different.

In this paper, we are interested in transport properties of mesoscopic diffusive wires, where quantum interferences lead to small deviations to Ohm's law. This is the so-called {\it weak-localisation regime}.
For classical transport, simple application of Ohm's law leads to the expression of the transmission coefficients, which amounts to classical addition of resistances and conductances (Kirchhoff). They are expressed in terms of the elements of a matrix which encodes the conductances of all the links of the network, defined below in Eq.\eqref{eq:DefM0}.

For a single wire of length $L$, it is well-known that the weak-localisation correction to the classical transmission coefficient can be written as~\cite{AkkMon07}
\begin{equation}
  \label{eq:WLwire}
  \Delta T  = - \frac{2}{L^2} \int_0^L \D x\,  P_c(x,x) 
  \:, 
\end{equation}
where $\Delta T=\mean{T}-T_\mathrm{class}$ is an average over disorder configurations. The so-called Cooperon $P_c(x,x)$ 
measures the contribution of interfering closed electronic diffusive trajectories.
We have shown that in a network of diffusive wires, this simple relation generalizes to~\cite{TexMon04}
\begin{equation}
  \label{eq:TexierMontambauxPRL2004}
  \Delta T_{\alpha\beta}  =
  \frac{2}{\xiloc} \sum_{i} 
  \frac{\partial T_{\alpha\beta}^\mathrm{class} }
       {\partial\, l_{i}}
  \int_{\mathrm{wire}\:(i)}\hspace{-0.5cm}\D x\, P_c(x,x)
  \:,
\end{equation}
where $i$ labels all the wires of the network. 
$\xiloc=\alpha_d N_c\ell_e$ is the localisation length~\footnote{Note that the perturbative approach is valid for $\min(L_\varphi,\mathrm{size})\ll\xiloc$.} 
in the infinitely long wire with $N_c$ conducting channels~\cite{Bee97}, $\ell_e$ the elastic mean free path and $\alpha_d$ a dimensionless parameter of order unity, which will be given below.

From the knowledge of the quantum corrections \eqref{eq:TexierMontambauxPRL2004}, we will show in this paper that the quantum correction to the classical four-terminal resistance is
\begin{equation}
  \label{eq:WL4TR}
  \Delta\mathcal{R}_{\alpha\beta,\mu\nu}  =
  \frac{2}{\xiloc} \sum_{i} 
  \frac{\partial \mathcal{R}^\mathrm{class}_{\alpha\beta,\mu\nu} }
       {\partial\, l_{i}}
  \int_{\mathrm{wire}\:(i)}\hspace{-0.5cm}\D x\, P_c(x,x)
  \:.
\end{equation}
This expression is quite simple since the weights attached to each wire have a simple interpretation~: they express the sensitivity of the classical four-terminal resistance when the resistance of this wire is modified.

Similarly we have found convenient expressions for the correlation functions of the transmission coefficients, from which one can obtain the correlation functions of the four-terminal conductances. Like the weak-localisation correction, these expressions involve contribution of all the wires, which are weighted by similar factors $\partial {\cal R}^{\text{class}}_{\alpha\beta,\mu\nu}/\partial \, l_i$,
Eqs.~(\ref{eq:4TRcorrel1},\ref{eq:4TRcorrel2},\ref{eq:4TRcorrel3},\ref{eq:4TRcorrel4}). These equations, which are, with Eq.~\eqref{eq:WL4TR}, the main results of the article, will be illustrated by several examples in simple devices.

\vspace{0.25cm}

Before going specifically to the analysis of quantum transport in networks of quasi one-dimensional weakly disordered wires, we close the section with some general remarks as the concept of four-terminal resistance (4TR), of which B\"uttiker has been one of the main promotors, has been extremely fruitful in mesoscopic physics. Let us mention few directions~:
\begin{itemize}
\item 
  In the early developments of the Landauer-B\"uttiker approach, the concept of four-terminal resistance has helped clarifying the question of contact resistance. 
  The concept of contact resistance is the ``mesoscopic version'' of the electric resistance appearing when the electronic fluid is injected from a macroscopic conductor into a small hole (known as Sharvin resistance \cite{Sha65}).
The role of contact resistances has been nicely explained in several papers of B\"uttiker \cite{ButImrLanPin85,But86a,But86,But88a} (see also chapter~5 of~\cite{Imr97}).

\item
  A fundamental aspect of quantum transport concerns the symmetry of transport coefficients~:
  symmetry with respect to current sources and voltage probes exchange, and symmetry with respect to the reversal of the magnetic field $\mathcal{B}$.
  Extending these ideas to coherent conductors, B\"uttiker has demonstrated the symmetry relation for the 4TRs~\cite{But86a,But88a}
\begin{equation}
  \label{eq:Buttiker88CasOns}
   \mathcal{R}_{\alpha\beta,\mu\nu}(-\mathcal{B})
  =\mathcal{R}_{\mu\nu,\alpha\beta}(\mathcal{B}) 
  \:.
\end{equation}
        
\item 
  The concept of 4TR provides an illuminating description of the integer quantum Hall effect \cite{But88} from the edge state picture introduced by Halperin \cite{Hal82}, as it allows to compute straightforwardly the longitudinal and Hall resistances (see also B\"uttiker's beautiful review article \cite{But91}).
  Furthermore, this framework permits to analyze in simple terms other more subbtle effects, like the scattering between edge states at opposite boundaries due to a constriction, leading to so-called anomalous Hall effect~\cite{KomHirSasHiy89,WeeWilHarBeeHouWilFoxHar89,But91}, or by impurities \cite{But89}, the description of transport in a Hall cross \cite{ForWasButKnoHon89}, etc.
  
\item 
  Another issue which has been put forward by B\"uttiker with others, and which will be central in the present article, concerns the \textit{nonlocal} nature of quantum transport, and the influence of voltage probes on the transport properties of a coherent conductor.
  Motivated by a set of experiments in multiconnected metallic wires \cite{BenUmbLaiWeb87,SkoManHowJacTenSto87} (see also Whasburn and Webb's review \cite{WasWeb86}), various authors have analyzed the role of voltage probes in devices made of disordered wires by various approaches~: Maekawa, Isawa and Ebisawa \cite{MaeIsaEbi87}, B\"uttiker \cite{But87}, Divincenzo , Kane and Lee \cite{KanLeeDiv88,DivKan88}, 
Chandrasekhar, Prober and Santhanam \cite{San87,San89,ChaSanPro91}, Hershfield and Ambegaokar \cite{HerAmb88,Her89}.
  A more general discussion of nonlocality of weak localisation in networks of metallic wires was made possible within the theory developed by us in Ref.~\cite{TexMon04}. Our formalism has allowed us to study how Altshuler-Aronov-Spivak oscillations of the magnetoconductance are affected by the network geometry \cite{DouRam85,PanChaRamGan85,DouRam86,FerAngRowGueBouTexMonMai04,TexMon05,SchMalMaiTexMonSamBau07,TexDelMon09} (this was beautifully demonstrated by earlier experiments in arrays of lithium rings by Bishop, Dolan and Licini \cite{DolLicBis86}), or the role of electronic interactions \cite{LudMir04,TexMon05b,Tex07b,TexMon07c,FerRowGueBouTexMon08,TexDelMon09,TreYevMarDelLer09,TreTexYevDelLer11,CapTexMonMaiWieSam13}.


\end{itemize}

The outline of the paper is as follows~:
in the next section we introduce a classical description of electronic transport in networks of metallic wires.
Section~\ref{sec:WL} is devoted to the analysis of the weak localisation correction, as a warm up exercice preparing the more complicate study of four-terminal resistance correlations presented in Section~\ref{sec:Fluc}.
The main formulae, Eqs.~\eqref{eq:WL4TR}, etc., are illustrated on simple cases.
Section~\ref{sec:Conclu} closes the paper with some concluding remarks.


\section{Classical transport in networks}

\subsection{Electrostatic potential}

\begin{figure}[!ht]
\centering
\includegraphics[width=0.4\textwidth]{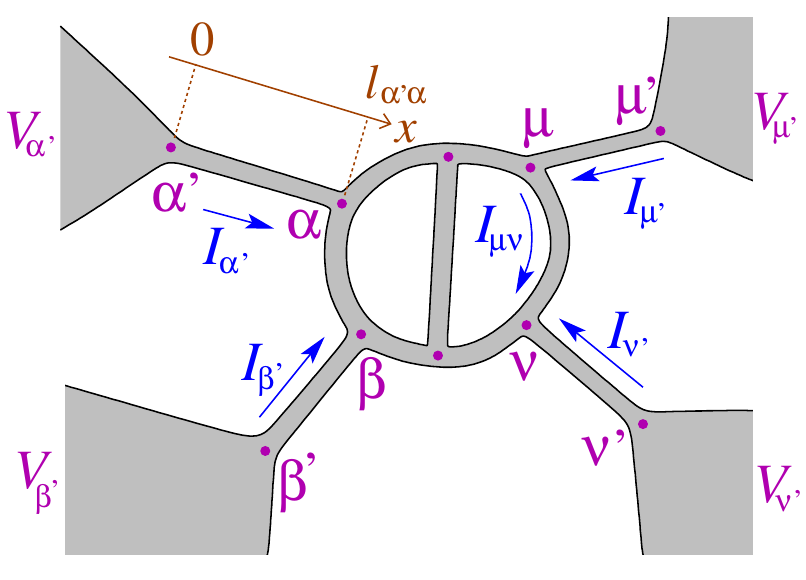}
\caption{\textit{A four-terminal mesoscopic network of metallic wires.
         Vertices are marked with magenta dots.
         Internal vertices are characterised by $\lambda_\mu=0$.
         Primed vertices correspond to reservoirs for which $\lambda_{\mu'}=\infty$.}}
\label{fig:net4t}
\end{figure}

A typical network is represented in Fig.~\ref{fig:net4t}. In this section, we introduce a \textit{specific notation in order to distinguish between internal vertices and vertices corresponding to reservoirs, by labelling these latter with a prime}.
Before going to the discussion of quantum transport, let us analyze the classical transport in the network. 
For this purpose we start by solving the Poisson equation $\Delta V(\vec{r})=0$ for the electrostatic potential $V(\vec{r})$ inside the network. 
Boundary conditions are $V(\vec{r})=V_{\alpha'}$ for $\vec{r}\in\mathrm{reservoir}\ \alpha'$.
Introducing the coordinate $x$ measuring the distance along the wire $\mu\nu$ (from $\mu$ to $\nu$), the potential varies linearly as $V(x)=V_\mu\,(1-x/l_{\mu\nu})+V_\nu\,x/l_{\mu\nu}$.
The (classical) current density is given by Fick's law $\vec{j}(\vec{r})=-eD\vec{\nabla}\delta n(\vec{r})$, where $D$ is the diffusion constant. 
The density in excess $\delta n$ is related to the potential through the effective (screened) Coulomb interaction $eV(\vec{r})=(1/\nu_0)\,\delta n(\vec{r})$, leading to $\vec{j}(\vec{r})=-\sigma_0\vec{\nabla}V(\vec{r})$, where $\sigma_0=e^2\nu_0D$ is the Drude conductivity and $\nu_0$ the density of states at Fermi energy.
The current in the wire $\mu\nu$ is given by \eqref{eq:Ohm}, hence current conservation at each ``internal'' vertex $\mu$, $\sum_{\nu\ \mathrm{neighbour\ of\ }\mu}I_{\mu\nu}=0$, may be rewritten as 
$\sum_\nu\big(\mathcal{M}_0\big)_{\mu\nu}V_\nu=0$, where the matrix is defined as
\begin{equation}
  \label{eq:DefM0}
  \left(\mathcal{M}_0\right)_{\mu\nu} 
  = \delta_{\mu\nu} \left( \lambda_\mu +\sum_\alpha \frac{a_{\mu\alpha}}{l_{\mu\alpha}}\right)
  - \frac{a_{\mu\nu}}{l_{\mu\nu}}
  \:.
\end{equation}
$a_{\mu\nu}$ is the adjacency matrix element, equals to $1$ if a wire connects the two vertices and $0$ otherwise~; thus it constraints the sum in \eqref{eq:DefM0} to run over vertices neighbours of $\mu$.
The parameters $\lambda_\mu$ have been introduced for convenience for the following and describe connection to reservoirs ($\lambda_\mu=0$ for an internal vertex and $\lambda_{\mu'}\to\infty$ for a vertex in a reservoir, cf. Fig.~\ref{fig:net4t}).
Up to a factor $\sw\sigma_0$, the matrix $\mathcal{M}_0$ thus simply gathers all the wire conductances  $\sigma_0\sw/l_{\mu\nu}$, where $\sw$ is the cross section of the wires.
If we split the vector gathering the electrostatic potentials at the vertices into two parts related to internal and external vertices,
$(V_\mathrm{in}|V_\mathrm{res})=(\cdots,V_\alpha,\cdots|\cdots,V_{\alpha'},\cdots)$, we may write 
$\big(\mathcal{M}_0\big)_\mathrm{in,in}V_\mathrm{in}
=-\big(\mathcal{M}_0\big)_\mathrm{in,res}V_\mathrm{res}$,
i.e.~\footnote{since $\lambda_{\alpha'}\to\infty$, the inverse of the matrix is simply the inverse of the block $\big(\mathcal{M}_0\big)_\mathrm{in,in}$ related to internal vertices.
  All other matrix elements are zero, $\big(\mathcal{M}_0^{-1}\big)_{\mu'\beta}=0$ for any reservoir $\mu'$.
} 
$V_\alpha = \sum_{\mathrm{res.}\:\beta'} \left(\mathcal{M}_0^{-1}\right)_{\alpha\beta}V_{\beta'}/l_{\beta\beta'}$ where the sum runs over the reservoirs.
It will be convenient for the following to introduce the notation $P_d(\alpha,\beta)=\left(\mathcal{M}_0^{-1}\right)_{\alpha\beta}$ which represents the so-called Diffuson, measured at the two vertices~:
\begin{equation}
  \label{eq:Potential0}
  V_\alpha = \sum_{\mathrm{res.}\:\beta'}
  P_d(\alpha,\beta)\,\frac{1}{l_{\beta\beta'}}\,V_{\beta'}
  \:.
\end{equation}
The Diffuson is solution of the diffusion equation 
\begin{equation}
  \label{eq:EqDiffuson}
  -\partial_x^2 P_d(x,x')=\delta(x-x')
\end{equation} 
with Dirichlet boundary conditions at the reservoirs~: $P_d(x,\alpha')=P_d(\alpha',x)=0$ for all reservoirs $\alpha'$ (for details, cf. Appendix of Ref.~\cite{TexDelMon09}).
Using the linearity of the Diffuson on the wires and that it vanishes at the reservoirs, we write
$P_d(\bullet,x)=(x/l_{\beta\beta'})\,P_d(\bullet,\beta)$ for $x\in\beta'\beta$ (the reservoir is at $x=0$). 
Hence, we can rewrite the coefficients in \eqref{eq:Potential0} as 
$P_d(\alpha,\beta)/l_{\beta\beta'}=\partial_{x'}P_d(\alpha,x')$, where $x'$ is any position in the wire $\beta'\beta$.
In the following we will prefer to write the slope of the Diffuson on the wire $\beta\beta'$ as 
$P_d(\alpha,\underline{\beta'})/\ell_e$, where $P_d(\bullet,\underline{\beta'})$ denotes that the argument is taken at a distance $\ell_e$ of the reservoir $\beta'$, $\ell_e$ being the elastic mean free path, i.e. the smallest length scale of the problem.
The Diffuson at a distance $\ell_e$ naturally appears in the diagrammatic calculation of the classical transport coefficients~\cite{TexMon04}.
The relation with this reference's notations will be clear by using below this expression for the Diffuson's slope.
Since the vertex $\alpha$ may be any point of the network, we may rewrite \eqref{eq:Potential0} as
\begin{equation}
 \label{eq:Potential}
 V(x) = \sum_{\mathrm{res.}\:\beta'}
 \frac{P_d(x,\underline{\beta'})}{\ell_e} \, V_{\beta'}
 \:;
\end{equation}
the expression \eqref{eq:Potential} is only valid when $x$ is at distance larger than $\ell_e$ from the reservoirs.

\subsection{Current distribution and generalised conductances}

Integrating the current density $j_x(x)=-\sigma_0\partial_xV(x)$ across the section $\sw$ of the wire $\mu\nu$, we obtain the current under the form~:
\begin{equation}
  \label{eq:CurrentInside}
  I_{\mu\nu}
  =  \sum_{\mathrm{res.}\:\beta'}
  \Gintern_{\mu\nu,\beta'}\,V_{\beta'}
\end{equation}
where we have introduced
\begin{equation}
  \label{eq:DefUpsilon}
  \Gintern_{\mathrm{wire\:}\mu\nu,\mathrm{res.}\:\beta'} = 
 -\frac{\sigma_0\sw}{\ell_e}\,\derivp{P_d(x,\underline{\beta'})}{x}
 \hspace{0.5cm}\mbox{for }
 x\in\:\mu\nu
 \:.
\end{equation}
The quantity $\Gintern_{\mu\nu,\beta'}$ is a generalised conductance matrix relating the \textit{external} potentials to the \textit{internal} currents. 
They obviously satisfy the symmetry property $\Gintern_{\mu\nu,\beta'} =-\Gintern_{\nu\mu,\beta'}$.
Although the physical interpretation was not provided in Ref.~\cite{TexMon04}, the explicit expression of \eqref{eq:DefUpsilon} in terms of the matrix $\mathcal{M}_0$ was given~:
\begin{equation}
  \label{eq:GinterneM0}
  \frac{\Gintern_{\mu\nu,\beta'}}{\sigma_0\sw}
  = \frac{\left(\mathcal{M}_0^{-1}\right)_{\beta\mu}
    - \left(\mathcal{M}_0^{-1}\right)_{\beta\nu}
    - \delta_{\mu\beta}\delta_{\nu\beta'}\,l_{\beta\beta'}}{l_{\beta\beta'}l_{\mu\nu}}
    \:.
\end{equation}

If we consider the case of a wire connected to a reservoir, $\mu\nu\to\alpha'\alpha$, Eq.~\eqref{eq:CurrentInside} coincides with the usual relation between currents and voltage in the terminals~:
\begin{equation}
  I_{\alpha'}\equiv I_{\alpha'\alpha}
  = \sum_{\mathrm{res.}\:\beta'} G_{\alpha'\beta'}^\mathrm{class}\, V_{\beta'}
  \:.
\end{equation}
The conductance matrix is obviously related to the generalised conductances by 
\begin{align}
  G_{\alpha'\beta'}^\mathrm{class} =\Gintern_{\alpha'\alpha,\beta'}
  \:.
\end{align}
The expression of the classical conductance matrix in terms of the matrix $\mathcal{M}_0$ may be deduced by setting $\mu\nu\to\alpha'\alpha$ in \eqref{eq:GinterneM0}~: we recover the expression of Ref.~\cite{TexMon04}
\begin{align}
  \label{eq:Gclassique}
  G_{\alpha'\beta'}^\mathrm{class} &= - \frac{2_se^2}{h}\alpha_dN_c\,
  \frac{P_d(\underline{\alpha'},\underline{\beta'})}{\ell_e}
  \nonumber\\
  & =- \frac{2_se^2}{h}\alpha_dN_c\ell_e\,
  \frac{\left(\mathcal{M}_0^{-1}\right)_{\alpha\beta}}{l_{\alpha\alpha'}l_{\beta\beta'}}
  \:.
\end{align}
We have used $\sigma_0\sw=(2_se^2/h)\alpha_dN_c\ell_e$ , where $2_s$ is the spin degeneracy, $N_c$ the number of conducting channels and $\alpha_d=V_d/V_{d-1}$ involves the volume of the $d$-dimensional sphere of unit radius (thus $\alpha_1=2$, $\alpha_2=\pi/2$ and $\alpha_3=4/3$).

\paragraph{Example}

\begin{figure}[!ht]
\centering
\includegraphics[width=0.3\textwidth]{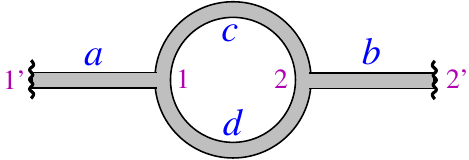}
\caption{\textit{A ring connected to two reservoirs.}}
\label{fig:Ring}
\end{figure}

As a simple illustration of Eq.~\eqref{eq:Gclassique}, we consider the ring of Fig.~\ref{fig:Ring}.
From the definition \eqref{eq:DefM0}, we write the internal part of the matrix $\mathcal{M}_0$ (i.e. the block related to vertices 1 and 2)~: 
\begin{equation} 
  \big(\mathcal{M}_0\big)_\mathrm{in,in}
  =\begin{pmatrix}
    1/l_a+1/l_{c\parallel d} & - 1/l_{c\parallel d} \\
    - 1/l_{c\parallel d} & 1/l_{c\parallel d} + 1/l_b
  \end{pmatrix}
  \:,
\end{equation}
where $1/l_{c\parallel d}=1/l_c+1/l_d$.
Eq.~\eqref{eq:Gclassique} leads to 
\begin{equation}
  \frac{1}{l_al_b}\left(\mathcal{M}_0^{-1}\right)_{12}
  =\frac{1}{l_a+l_{c\parallel d}+l_b}
\end{equation}
giving the expected conductance.

\subsection{Four-terminal resistances}

Using \eqref{eq:Buttiker1986}, the 4TRs can be deduced from the conductance matrix \eqref{eq:Gclassique} (see also the discussion in \ref{app:CRFT}).
For simple enough networks, the determination of the resistances is however more simple than that of the conductance matrix and does not require the knowledge of the latter, as the simple example analyzed just before has shown.


\section{Weak localisation}
\label{sec:WL}

Weak localisation is a small quantum correction to transport coefficients originating from quantum interferences between time reversed electronic trajectories~\cite{AltLee88,AkkMon07}.
The main interest in this small quantum correction to transport coefficients is that it gives a measure of the phase coherence length $L_\varphi$, the fundamental characterisitc length scale which sets the boundary between quantum and classical physics. 
It is worth stressing that there is no intrinsic definition of $L_\varphi$, which can only be obtained by extracting a characteristic length scale from the analysis of a physical quantity sensitive to quantum interference, such as weak localisation.
A precise experimental determination of $L_\varphi$ thus requires a perfect knowledge of the functional form of the transport coefficients as a function of the various length scales, the magnetic field, etc.

\subsection{Conductance matrix}
\label{subsec:3.1}

The weak localisation correction to the conductance matrix elements $G_{\alpha'\beta'}=-(2_se^2/h)\,T_{\alpha'\beta'}$ is given by~\cite{TexMon04}
\begin{equation}
  \label{eq:WLgeneral}
  \Delta T_{\alpha'\beta'} = 
  \frac{2}{\ell_e^2}\int_\mathrm{Network}\hspace{-0.5cm}\D x \,
  \derivp{P_d(\underline{\alpha'},x)}{x}\,P_c(x,x)\,
  \derivp{P_d(x,\underline{\beta'})}{x}
  \:,
\end{equation}
where $P_c(x,x')$ is the Cooperon, solution of 
\begin{equation}
  \label{eq:EqCooperon}
  \left({1}/{L_\varphi^2} - D_x^2\right)P_c(x,x') = \delta(x-x') 
  \:,
\end{equation}
$D_x=\partial_x-2\I eA(x)$ being the covariant derivative and $L_\varphi$ the phase coherence length.
Few remarks~:
\begin{itemize}
\item
  The effect of the magnetic field is twofold~:
  (\textit{i}) in the presence of loops in the network, the Cooperon depends on the magnetic fluxes, which leads to Altshuler-Aronov-Spivak oscillations~\cite{AltAroSpi81,AltAroSpiShaSha82,AroSha87} (see also \cite{TexMon05,TexDelMon09}).
  (\textit{ii}) The penetration of the magnetic field in a narrow wire of width $w$ can be accounted for through the substitution $1/L_\varphi^2\to1/L_\varphi^2+1/L_{\mathcal{B}}^2$, where the magnetic length is 
$L_\mathcal{B}=[\sqrt{3}/(2\pi)]\,\phi_0/(\mathcal{B}w)$ and $\phi_0=h/e$ the quantum flux~\cite{AltAro81}.

\item
The expression \eqref{eq:WLgeneral} is of great generality~: it is not only valid for a system made of quasi-1D wires (network) but only assume that the contacts have a quasi-1D geometry. In such a more general situation, the derivatives should simply be replaced by gradients 
$\partial_xP_d(\underline{\alpha'},x)\partial_xP_d(x,\underline{\beta'})\to
\vec{\nabla}P_d(\underline{\alpha'},\vec{r})\cdot\vec{\nabla}P_d(\vec{r},\underline{\beta'})$.

\item
  Expression \eqref{eq:WLgeneral} is valid for $\alpha'\neq\beta'$. 
  The direct diagrammatic calculation of the reflection probability is more difficult and involves a description of the matching between the metallic system and the contacts which goes beyond the derivation of \eqref{eq:WLgeneral}~\cite{HasStoBar94,TexMon04}.
However the WL correction to the diagonal conductance matrix elements can always be deduced by using current conservation $\sum_\alpha G_\ab=0$.

\item
Interestingly, we see that the contribution of each wire is weighted by the ``internal conductances'' introduced above~:
\begin{align}
  \Delta T_{\alpha'\beta'} = 
  \frac{2}{(\sw\sigma_0)^2} \sum_{(\mu\nu)}
  &\Gintern_{\mu\nu,\alpha'} \Gintern_{\mu\nu,\beta'} 
  \\\nonumber
  &\times\int_{\mathrm{wire}\:(\mu\nu)}\hspace{-0.5cm}\D x\, P_c(x,x)
  \:.
\end{align}
This expression shows that a uniform integration of the Cooperon in the network is possible only if the distribution of the classical currents in the wires is uniform.
\end{itemize}

More conveniently, we showed in Ref.~\cite{TexMon04} that these weights are related to the derivatives of the classical conductance matrix, Eq.~\eqref{eq:TexierMontambauxPRL2004}, or equivalently~:
\begin{align}
  \label{eq:TexMonPRL04}
  \Delta G_{\alpha\beta} =
  \frac{2}{\xiloc} \sum_{i} 
  \frac{\partial G^\mathrm{class}_{\alpha\beta}}{\partial\,l_{i}}
  \int_{\mathrm{wire}\:(i)}\hspace{-0.5cm}\D x\, P_c(x,x)
  \:,
\end{align}
where $\xiloc=\alpha_d N_c\ell_e$ is the localisation length in the infinitely long wire.
\textit{From now, we will drop the prime on the vertices connected to reservoirs, as there will be no possible confusion below}.

\subsection{Nonlocality leading to positive WL correction}

\begin{figure}[!ht]
\begin{center}
\includegraphics[width=0.2\textwidth]{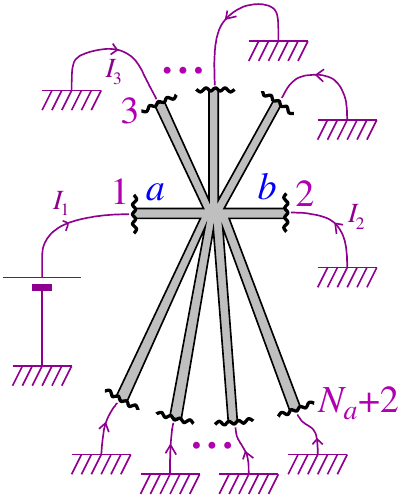}
\end{center}
\caption{\it For a sufficient large number $N_a$ of long wires, the WL correction to the transmission $\Delta T_{12}$ may be become positive~\cite{TexMon04}.}
\label{fig:netPosWL}
\end{figure}

An interesting consequence of the nature of the weighting factors was pointed out in Ref.~\cite{TexMon04}~: 
since the weight $\partial G^\mathrm{class}_{\alpha\beta}/\partial l_i$ may change in sign for certain wires, the WL correction to some transmission coefficient may become positive. Such an example is shown on Fig.~\ref{fig:netPosWL}, which has
\begin{equation}
  \Delta T_{12} \simeq \frac{1}{3}
  \left( -1 + N_a \frac{l_{a\parallel b}}{l_a+l_b} \right)
  \:,
\end{equation}
valid for $N_a$ long wires ($\gg l_a,\,l_b$). 
This WL correction may become positive for sufficiently large~$N_a$, as a striking illustration of the nonlocality of quantum transport.

\subsection{Effect of nonlocality on the four-terminal resistances}

We have demonstrated that a relation similar to \eqref{eq:TexMonPRL04} holds for the 4TRs, Eq.~\eqref{eq:WL4TR} (cf.~\ref{appendix:proof}).
We illustrate this formula by considering the resistances $\mathcal{R}_{12,12}$ and $\mathcal{R}_{12,34}$ characterizing the network of Fig.~\ref{fig:Wire4TR}.

A remarkable consequence of quantum nonlocality is the possibility of large WL correction to the conductance, induced by the presence of long 1D contacts~\cite{San87}.
To illustrate this idea, we analyze the WL correction to the two resistances $\mathcal{R}_{12,12}$ and $\mathcal{R}_{12,34}$ for the network of Fig.~\ref{fig:Wire4TR}.

\textit{In the following we will express the resistances in units of $h/(2_se^2)$.}

\subsubsection{Reminder~: two-terminal configuration}

At this stage it is useful to recall the well-known result for WL correction to the conductance (or resistance) in a two-terminal measurement (a wire of length $L$ between two large contacts). In this case the dimensionless resistance is simply $\mathcal{R}_\mathrm{class}=L/\xiloc$ and the Cooperon satisfying Dirichlet boundary condition reads $P_c(x,x')=L_\varphi\sinh(x_</L_\varphi)\sinh((L-x_>)/L_\varphi)/\sinh(L/L_\varphi)$, where $x_<=\min(x,x')$ and $x_>=\max(x,x')$. The application of \eqref{eq:WLwire} gives the result of Ref.~\cite{AltAroZyu84}~:
\begin{equation}
  \label{eq:AltAroZyu84}
  \frac{\Delta\mathcal{R}}{\mathcal{R}_\mathrm{class}^2}
  =\frac{L_\varphi}{L}
  \left[\coth(L/L_\varphi)-\frac{L_\varphi}{L}\right]
  \simeq\frac{L_\varphi}{L}-\left(\frac{L_\varphi}{L}\right)^2 
  \:,
\end{equation}
where the expansion corresponds to $L_\varphi\ll L$.
The dominant term corresponds to the value of the Cooperon in bulk (i.e. inside a wire, at distance $\gg L_\varphi$ from the boundaries)~; the correction is explained by the depletion of the Cooperon which vanishes at the boundaries.
In the coherent limit ($L_\varphi\to\infty$), one get the universal WL correction 
$\Delta\mathcal{R}/\mathcal{R}_\mathrm{class}^2\simeq1/3$.
Note that \eqref{eq:AltAroZyu84} may be obtained more directly by using the spectral determinant approach \cite{AkkComDesMonTex00,ComDesTex05}.

\subsubsection{Two-terminal resistance $\mathcal{R}_{12,12}$ for multiconnected wire}

We consider the device of Fig.~\ref{fig:Wire4TR} and consider first the two-terminal resistance. Due to nonlocality, although the two arms $d$ and $f$ play no role for the classical transport, $\mathcal{R}_{12,12}^\mathrm{class}=(l_a+l_b+l_c)/\xiloc$, they influence the quantum contributions to the transport coefficients.  
Eq.~(\ref{eq:WL4TR}) implies that integration of the Cooperon runs over the three wires $(a)$, $(b)$ and $(c)$~:
\begin{equation}
  \frac{\Delta\mathcal{R}_{12,12}}{(\mathcal{R}_{12,12}^\mathrm{class})^2}
  =\frac2{L^2}
  \left(\int_{(a)}+\int_{(b)}+\int_{(c)}\right)\D x\,P_c(x,x)
  \:.
\end{equation}
where $L=l_a+l_b+l_c$.
The explicit expression of the integral $\int_{(i)}\D x\,P_c(x,x)$ over a wire in terms of the network properties can be found in Ref.~\cite{TexMon04}. 
Below, we only analyze limiting behaviours.

\paragraph{Weakly coherent limit $L_\varphi\ll l_a,\, l_b,\, l_c,\, l_d,\, l_f$ ---}
We get in this case
\begin{equation}
  \label{eq:WLR1212}
  \frac{\Delta\mathcal{R}_{12,12}}{(\mathcal{R}_{12,12}^\mathrm{class})^2}
  \simeq\frac{L_\varphi}{L}-\frac23\left(\frac{L_\varphi}{L}\right)^2
  \:,
\end{equation}
up to exponentially small corrections. 
If we compare this expression with \eqref{eq:AltAroZyu84}, we remark that the presence of the two long wire $(d)$ and $(f)$ only affects the subleading term of the WL correction. 
The difference correction is explained by the depletion of the Cooperon at the vertices, which behaves as $P_c(x,x)\simeq L_\varphi/2-(L_\varphi/6)\EXP{-2x/L_\varphi}$ at distance $x$ of the vertex. 
The correction to the bulk result is 
$-4\times(2/L^2)\int_0^\infty\D x\,(L_\varphi/6)\EXP{-2x/L_\varphi}=-(2/3)(L_\varphi/L)^2$.

\paragraph{Fully coherent limit ($L_\varphi\to\infty$) ---}
We only discuss two limiting cases as the general expression is rather cumbersome. 
When the two wires $(d)$ and $(f)$ are very long, $l_d,\,l_f\gg l_a,\,l_b,\,l_c$, they play no role and we recover the universal result for the two-terminal wire [limit $L_\varphi\to\infty$ of Eq.~\eqref{eq:AltAroZyu84}]
$
{\Delta\mathcal{R}_{12,12}}/{(\mathcal{R}_{12,12}^\mathrm{class})^2}
\simeq1/3
$.

In the opposite limit $l_d,\,l_f\ll l_a,\,l_b,\,l_c$, the phase coherence is broken at the level of the vertices due to the vicinity of the large contacts. We get the result
$
{\Delta\mathcal{R}_{12,12}}/{(\mathcal{R}_{12,12}^\mathrm{class})^2}
\simeq(1/3)(l_a^2+l_b^2+l_c^2)/(l_a+l_b+l_c)^2
$,
which corresponds to the addition of resistances 
$\Delta\mathcal{R}_{12,12}\simeq\Delta R_a+\Delta R_b+\Delta R_c$ for three independent wires with $\Delta R_a=(1/3)(l_a/\xiloc)^2$, etc.

Varying the length of the wires $(d)$ and $(f)$ hence allows to cross over between the fully quantum regime where the wires $(a)+(b)+(c)$ should be considered as a whole and the regime where the three wires are independent and their resistances may be added according to the classical Kirchhoff law.
This is an illustration of the idea introduced by B\"uttiker to describe dephasing in a fully coherent system by introducing fictitious voltage probes~\cite{But86}.

\subsubsection{Four-terminal resistance $\mathcal{R}_{12,34}$~: nonlocality leading to large WL correction}

Classical resistance reads $\mathcal{R}_{12,34}^\mathrm{class}=l_b/\xiloc$ which immediately  shows that the WL correction is given by an integral of the Cooperon over the wire $(b)$ only~:
\begin{equation}
  \label{eq:IntR1234}
  \frac{\Delta\mathcal{R}_{12,34}}{(\mathcal{R}_{12,34}^\mathrm{class})^2}
  =\frac2{l_b^2}\int_{\mathrm{wire}\:(b)}\hspace{-0.5cm}\D x\,P_c(x,x)
  \:.
\end{equation}

\begin{figure}[!ht]
\begin{center}
\includegraphics[width=0.4\textwidth]{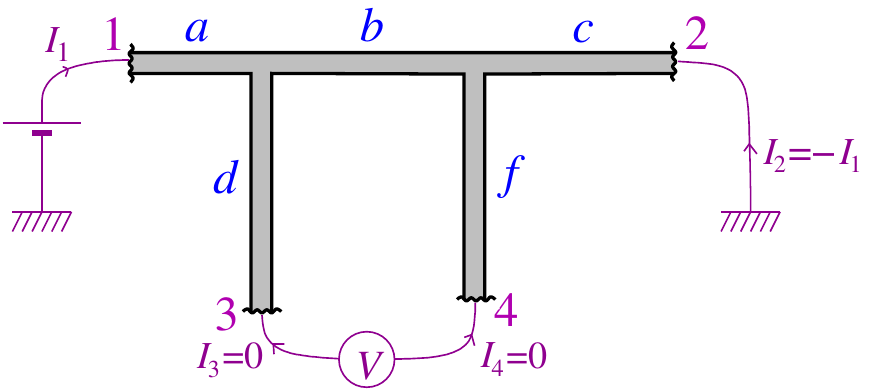}
\end{center}
\caption{\it A four-terminal device. When the central wire $(b)$ is shorter than the phase coherence length and the connecting wires, the WL correction $\Delta\mathcal{R}_{12,34}/(\mathcal{R}_{12,34}^\mathrm{class})^2$ is large.}
\label{fig:Wire4TR}
\end{figure}

\paragraph{Incoherent connecting wires $L_\varphi\ll l_a,\,l_c,\,l_d,\,l_f$ ---}

Using again the expression of the integral of the Cooperon integrated in a wire~\cite{TexMon04}, we obtain the explicit expression
\begin{align}
&\frac{\Delta\mathcal{R}_{12,34}}{(\mathcal{R}_{12,34}^\mathrm{class})^2}
\simeq
\frac{-1+\sqrt\gamma l_b\coth\sqrt\gamma l_b}{\gamma l_b^2}
\\\nonumber
&+\frac{1}{\gamma l_b^2}\frac2{4\coth\sqrt\gamma l_b+5}
\bigg[
  \frac{-1+\sqrt\gamma l_b\coth\sqrt\gamma l_b}{\sinh^2\sqrt\gamma l_b}
\\\nonumber
& +  
  \Big(\coth\sqrt\gamma l_b+2\Big)
  \left( \coth\sqrt\gamma l_b
        -\frac{\sqrt\gamma l_b}{\sinh^2\sqrt\gamma l_b}
  \right)
\bigg]
\end{align}
where $\gamma=1/L_\varphi^2$.
The first term is the result for an isolated wire of length $l_b$, Eq.~\eqref{eq:AltAroZyu84}. 
The  second term originates from the non vanishing value of the Cooperon at the two  vertices, {\it i.e.}  we can interpret this term as coming from the modification of the boundary conditions for the wire $(b)$ induced by  the presence of the connecting wires $(a)$, $(c)$, $(d)$ and $(f)$.
After a little bit of algebra, we obtain~:~\footnote{
  This result was obtained by Santhanam~\cite{San87}, although this paper does not provide a detailed discussion on how the Cooperon must be integrated in a complex geometry.}
\begin{equation}
  \frac{\Delta\mathcal{R}_{12,34}}{(\mathcal{R}_{12,34}^\mathrm{class})^2}
  \simeq\frac{L_\varphi}{l_b}
  \frac{5\coth(l_b/L_\varphi)+4-3L_\varphi/l_b}{4\coth(l_b/L_\varphi)+5}
  \:.
\end{equation}

\vspace{0.125cm}

\noindent$\bullet$ For a long wire $L_\varphi\ll l_b$ we get the small correction
\begin{equation}
  \label{wlR1234}
  \frac{\Delta\mathcal{R}_{12,34}}{(\mathcal{R}_{12,34}^\mathrm{class})^2}
  \simeq\frac{L_\varphi}{l_b}-\frac13\left(\frac{L_\varphi}{l_b}\right)^2
  \:,
  \end{equation}
up to exponentially small corrections.
The dominant term $L_\varphi/l_b\ll1$, coincides with the two-terminal measurement for the wire of  length $l_b$, Eq.~\eqref{eq:AltAroZyu84}. 
As for $\Delta\mathcal{R}_{12,12}$ the presence of the connecting  wires manifests itself only through the coefficient of the subleading term~$(L_\varphi/l_b)^2$, cf. \eqref{eq:AltAroZyu84} or \eqref{eq:WLR1212}. 
The subleading correction is half of the one obtained for $\Delta\mathcal{R}_{12,12}$, Eq.~\eqref{eq:WLR1212}, as only the depletion of the Cooperon at the two boundaries of the wire $(b)$ contributes.

\vspace{0.125cm}

\noindent$\bullet$ For a short (coherent) wire $l_b\ll L_\varphi$, quite remarkably, we obtain a \textit{large} WL correction to the four-terminal conductance
\begin{equation}
  \label{largeWL1}
  \frac{\Delta\mathcal{R}_{12,34}}{(\mathcal{R}_{12,34}^\mathrm{class})^2}
  \simeq\frac12\frac{L_\varphi}{l_b}\gg1
  \:.
\end{equation}
In this case the presence of the connecting wires strongly affects the Cooperon inside the wire, which is the reason for the large WL correction. 
In a wire connected to two large reservoirs, the size of the electronic trajectories contributing to the WL is bounded by the length of the wire, which leads to a saturation of $\Delta g$ as $L_\varphi\to\infty$. On the contrary, in the four-terminal configuration, the electronic trajectories can explore the connecting wires $(a)$, $(c)$, $(d)$ and $(f)$ on large scales compared to $l_b$, which is the physical origin for the large WL.

Eq.~\eqref{largeWL1} characterizes the WL correction to the conductance, in a four probe configuration. Although the correction to the conductance may be large, we remark that the \textit{relative} correction, 
\begin{equation}
  \frac{\Delta\mathcal{R}_{12,34}}{\mathcal{R}_{12,34}^\mathrm{class}}
  \simeq \frac{L_\varphi}{2\xiloc} \ll 1
\end{equation}
is always small, as the validity of the perturbative treatment requires  $L_\varphi\ll\xiloc$.

\begin{figure}[!ht]
\begin{center}
\includegraphics[width=0.3\textwidth]{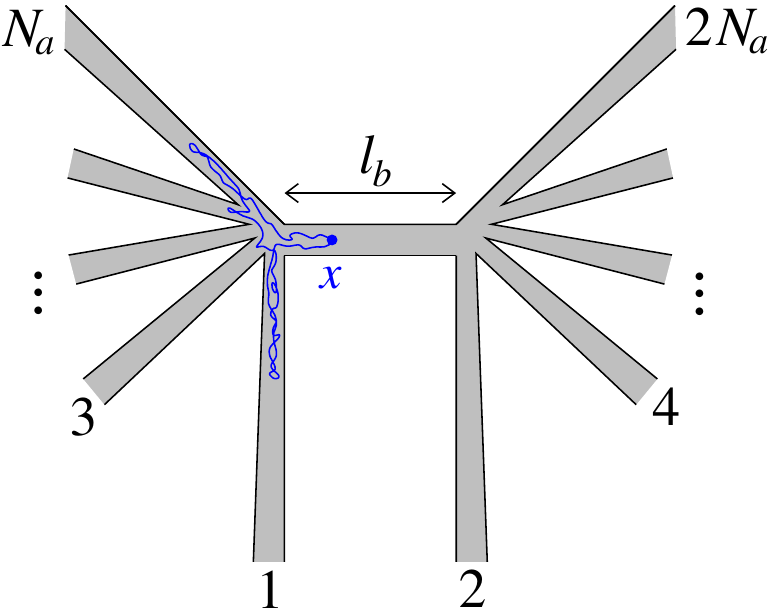}
\end{center}
\caption{\it A wire of length $l_b$ connected to $2N_a$ reservoirs by long wires.
  WL correction ${\Delta\mathcal{R}_{12,34}}/{(\mathcal{R}_{12,34}^\mathrm{class})^2}$ is controlled by electronic trajectories starting from wire $(b)$ which may explore the wires over long distances compared to $l_b$.}
\label{fig:MultiTerm}
\end{figure}

In order to better understand the difference between the two results \eqref{wlR1234} and \eqref{largeWL1}, it is instructive to consider the multiterminal network of Fig.~\ref{fig:MultiTerm}. The WL correction to  $\mathcal{R}_{12,34}$ is given by \eqref{eq:IntR1234} as well.
For the calculation, the key point is that, at a vertex $x$ from which issue $m_x$ long wires (longer than $L_\varphi$), the value of the Cooperon is $P_c(x,x)\simeq L_\varphi/m_x$ (cf. Ref.~\cite{TexDelMon09} and \ref{appendix:Star}).
In the weakly coherent wire limit ($L_\varphi\ll l_b$), the Cooperon inside the wire $(b)$ is $P_c(x,x)\simeq L_\varphi/2$, except at a distance $\lesssim L_\varphi$ from the vertices. Integration of the Cooperon leads to 
${\Delta\mathcal{R}_{12,34}}/{(\mathcal{R}_{12,34}^\mathrm{class})^2}\simeq L_\varphi/l_b\ll1$, similar to Eq.~\eqref{wlR1234}.
In the coherent limit ($L_\varphi\ll l_b$), with connecting wires still longer than $L_\varphi$, the Cooperon is almost uniform inside the wire, with a value $P_c(x,x)\simeq L_\varphi/(2N_a)$, where  $2N_a$ is the effective coordination number. We get in this case the large WL correction 
${\Delta\mathcal{R}_{12,34}}/{(\mathcal{R}_{12,34}^\mathrm{class})^2}\simeq L_\varphi/(N_al_b)$, which reduces to \eqref{largeWL1} for $N_a=2$.
This argument provides the interpretation of the factor $1/2$ in Eq.~\eqref{largeWL1}.

\paragraph{Fully coherent limit $L_\varphi\to\infty$ ---}
Coming back to the simple network of Fig.~\ref{fig:Wire4TR}, it is also interesting to consider the fully coherent limit. Using again the expression of the integral $\int_{(b)}\D x\,P_c(x,x)$ given in \cite{TexMon04}, some algebra leads to
\begin{equation}
  \frac{\Delta\mathcal{R}_{12,34}}{(\mathcal{R}_{12,34}^\mathrm{class})^2} 
  =
  \frac13 +\frac23\,
  \frac{ l_{a\parallel d} + l_{c\parallel f} 
         + 3{l_{a\parallel d}\,l_{c\parallel f}}/{l_b}
       }
       {l_{a\parallel d}+l_{c\parallel f}+l_b}
  \:,
\end{equation}
where $1/l_{a\parallel d}=1/l_a+1/l_d$, etc.
For short connecting wire $l_a,\,l_c,\,l_d,\,l_f\ll l_b$,
we recover the well known universal result $\Delta\mathcal{R}_{12,34}/(\mathcal{R}_{12,34}^\mathrm{class})^2\simeq1/3$ corresponding to a coherent wire between two large contacts, as expected.
In the other limit, $l_b\ll l_a=l_c=l_d=l_f$, we obtain a large correction
\begin{equation}
 \frac{ \Delta\mathcal{R}_{12,34}}{(\mathcal{R}_{12,34}^\mathrm{class})^2}\simeq 
 \frac12\frac{l_a}{l_b}
 \gg1 
\end{equation}
similar to \eqref{largeWL1} in which $L_\varphi\to l_a$ (i.e. the cutoff limiting the trajectories exploring the connecting wires is not the phase coherence length but the distance $l_a$ to the reservoirs).

\vspace{0.25cm}

The possibility for large WL correction was pointed out in Ref.~\cite{DouRam86} in the rather academic situation of an isolated wire.~\footnote{
  It is also well-known that the WL correction to the conductance is large in the two-dimensional situation, in a plane \cite{AkkMon07} leading to $\Delta g\simeq-(1/\pi)\ln(L/\ell_e)$ where $L$ is the size of the plane.
  However, the relative correction is small,  $\Delta g/g\simeq-2\ln(L/\ell_e)/(\pi k_F\ell_e)\ll1$, where $k_F$ is the Fermi wave vector. 
  The same logarithmic behaviour is obtained for large planar networks (square grids, honeycomb lattice, etc.) leading to $\Delta g\simeq-(1/\pi)\ln(L/a)$, where $a$ is the length of each wire \cite{TexDelMon09}.
} 
A more precise discussion was provided by Santhanam \cite{San87} for the case we have considered here.
As we already mentioned, this has the same origin as the large voltage fluctuations due to long coherent excursions of charge carriers in the voltage probes emphasized by B\"uttiker~\cite{But87} (and also in Ref.~\cite{BenUmbLaiWeb87}).
Although the observation of large WL corrections has not been reported so far, to the best of our knowledge, large resistance fluctuations have been observed in several  experiments~\cite{BenUmbLaiWeb87,SkoManHowJacTenSto87}, with the same physical origin.


\section{Fluctuations and correlations}
\label{sec:Fluc}

Mesoscopic (interference) phenomenon are more pronounced when the system size is reduced down to a size comparable to the phase coherence length $L_\varphi$~: the quantum contribution to the dimensionless conductance $\Delta g=g-g_\mathrm{class}$ of a fully coherent conductor presents fluctuations $\delta g=g-\mean{g}\sim1$ of the same order than the average $\mean{\Delta g}\sim1$ (the WL). For this reason, the characterization of conductance fluctuations/correlations has attracted considerable attention both  and experimentally~\cite{UmbWasLaiWeb84,WebWasUmbLai85,WasUmbLaiWeb85,UmbHaeLaiWasWeb86} and theoretically~\cite{AltShk86,LeeStoFuk87,ZyuSpi87} 
(for reviews, see \cite{SpiZyu91,WasWeb86,AkkMon07}).
The experiments are usually performed in the four-terminal configuration, which has brought the question of the role of the voltage probes~\cite{BenUmbLaiWeb87,SkoManHowJacTenSto87}.
Devices similar to the wire of Fig.~\ref{fig:Wire4TR} were studied in these experiments.
The nonlocal nature of quantum transport is particularly striking by considering the symmetric and antisymmetric parts of the resistance 
\begin{align}
  \label{defRSRA}
  \mathcal{R}_{S} &= \frac12\left(\mathcal{R}_{12,34}+\mathcal{R}_{34,12}\right)\\
  \mathcal{R}_{A} &= \frac12\left(\mathcal{R}_{12,34}-\mathcal{R}_{34,12}\right)
  \:,
\end{align}
which were shown to present different behaviours as a function of the ratio $l_b/L_\varphi$ (symmetrisation is done with respect to exchange of current and voltage probes or, thanks to Eq.~\eqref{eq:Buttiker88CasOns}, to magnetic field reversal).
Whereas $\mathcal{R}_{A}$ is a relatively flat function of $l_b/L_\varphi$, the symmetric resistance presents a clear crossover at $L_\varphi\sim l_b$, see Fig.~\ref{fig:benoit} (note that in a weakly disordered metal with a small enough magnetic field, we can ignore the classical magnetoresistance caused by the Lorentz force. As a consequence $\mathcal{R}_{A}^\mathrm{class}=0$).

The study of nonlocality of voltage fluctuations and/or transmission probabilities in multiterminal devices was considered theoretically by Maekawa \textit{et al.}~\cite{MaeIsaEbi87} and B\"uttiker \cite{But87} (for a three terminal device) by different approaches. The importance of long range potential correlations was later emphasized by Kane, Serota and Lee \cite{KanSerLee88}, which has led to reconsider the study of voltage fluctuations in an illuminating paper of Kane, Lee and DiVincenzo \cite{KanLeeDiv88}, and also in Refs.~\cite{HerAmb88,Her89} (note also the numerical study \cite{BarStoDiv88}).
Finally we point out few theoretical works on the ring configuration \cite{IsaEbiMae86,DivKan88} relevant for the experiments aforementioned, and specifically studied in Ref.~\cite{KurChaChiCha92}.

In the following we derive formulae analogous to (\ref{eq:TexierMontambauxPRL2004},\ref{eq:WL4TR}) for the correlations of transmissions and the correlations of four-terminal resistances
\begin{equation}
  \label{eq:DefCorrelator}
  \smean{
    \delta\mathcal{R}_{\alpha\beta,\mu\nu}(\mathcal{B})\,
    \delta\mathcal{R}_{\alpha'\beta',\mu'\nu'}(\mathcal{B}')
  }
  \:.
\end{equation}
We will apply our results to the analysis of the resistances $\mathcal{R}_{S}$ and $\mathcal{R}_{A}$ characterizing the four-terminal wire of Fig.~\ref{fig:Wire4TR}. 
We will show that our formalism allows to recover the results of Refs.~\cite{KanLeeDiv88,HerAmb88,Her89} straightforwardly.

\subsection{Conductance correlations}
\label{subsec:4.1}

Expression of the conductivity correlations in simple geometries can be found at several places~\cite{AltShk86,AkkMon07}.
In networks, the correlations of transmission coefficients 
$\smean{\delta T_{\alpha\beta}(\mathcal{B})\,\delta T_{\mu\nu}(\mathcal{B}')}$ (i.e. conductance matrix elements) are given by four contributions~\cite{TexMonAkk07}~:
\begin{align}
  \label{eq:Tucf1}
  &\smean{{T}_{\alpha\beta}{T}_{\mu\nu}}^{(1)} = 
  \frac{4}{\ell_e^4}
  \int\D\omega\,\delta_T(\omega)
  \int_{\rm Network}\hspace{-0.5cm}\D x\D x'\,
  \nonumber\\
  &\hspace{0.75cm}
  \derivp{P_d(\underline{\alpha},x) }{x} 
  \derivp{ P_d(\underline{\mu},x) }{x} \,
  P_\omega^\mathrm{(d)}(x,x')P_{-\omega}^\mathrm{(d)}(x,x')
  \nonumber\\
  &\hspace{0.25cm}
  \times 
  \derivp{P_d(x',\underline{\beta})}{x'} 
  \derivp{ P_d(x',\underline{\nu})}{x'}
  \\
\label{eq:Tucf2}
  &\smean{{T}_{\alpha\beta}{T}_{\mu\nu}}^{(2)} =
  \frac{4}{\ell_e^4}
  \int\D\omega\,\delta_T(\omega)
  \int_{\rm Network}\hspace{-0.5cm}\D x\D x'\,
  \nonumber\\
  &\hspace{0.75cm}
  \derivp{ P_d(\underline{\alpha},x)}{x}
  \derivp{ P_d(x,\underline{\nu})}{x}\,
  P_\omega^\mathrm{(c)}(x,x')P_{-\omega}^\mathrm{(c)}(x',x)
  \nonumber\\
  &\hspace{0.25cm}
  \times 
  \derivp{ P_d(\underline{\mu},x')}{x'}
  \derivp{ P_d(x',\underline{\beta})}{x'}
  \\
\label{eq:Tucf3}
  &\smean{{T}_{\alpha\beta}{T}_{\mu\nu}}^{(3)} =
  \frac{2}{\ell_e^4}
  \int\D\omega\,\delta_T(\omega)
  \int_{\rm Network}\hspace{-0.5cm}\D x\D x'\,
  \nonumber\\
  &\hspace{0.25cm}
  \derivp{ P_d(\underline{\alpha},x)}{x}
  \derivp{ P_d(x,\underline{\beta})}{x}\,
  \re\left[P_\omega^\mathrm{(d)}(x,x')P_\omega^\mathrm{(d)}(x',x)\right]
  \nonumber\\
  &\hspace{0.25cm}
  \times 
  \derivp{ P_d(\underline{\mu},x')}{x'}
  \derivp{ P_d(x',\underline{\nu})}{x'}
  \\
\label{eq:Tucf4}
  &\smean{{T}_{\alpha\beta}{T}_{\mu\nu}}^{(4)} =
  \mbox{ same as }  \smean{{T}_{\alpha\beta}{T}_{\mu\nu}}^{(3)}  
  \\ \nonumber
  & \hspace{2.5cm}\mbox{ with }
  P_\omega^\mathrm{(d)}\to P_\omega^\mathrm{(c)}
  \:,
\end{align}
where we used the same notation as before, $P_{d}(\underline{\alpha},x)$, in order to designate the Diffuson measured at a distance $\ell_e$ of the vertex $\alpha$.
The function $\delta_T(\omega)$ is a normalised function~\footnote{
  Its precise form for is
  $\delta_T(\omega)=F(\omega/2T)/2T$ with  
  $F(x)=(x\coth x-1)/\sinh^2x$.
} of width $\Delta\omega\sim T$ with $\delta_T(0)=1/(6T)$.
Several remarks~:
\begin{itemize}
\item 
  As for the WL, note that these expressions are of great generality, and not only valid for networks of quasi-1D wires~; they only assume contacts of quasi-1D nature.
  For a more general situation, one has to replace the derivatives as 
  $\partial_xP_d(\underline{\alpha},x)\partial_xP_d(\underline{\mu},x)
  \to
  \vec{\nabla}P_d(\underline{\alpha},\vec{r})\cdot\vec{\nabla}P_d(\underline{\mu},\vec{r})$.
  
\item 
  The Diffuson $P_d(x,x')$ connecting the contacts to the bulk, and providing the weights to attribute to each wire, obeys the classical diffusion equation \eqref{eq:EqDiffuson}.
  The Cooperon and Diffuson $P_\omega^\mathrm{(d,c)}(x,x')$, which describe phase coherent properties, are solutions of the diffusion equation
\begin{equation}
  \label{eq:EqDiffCoopFreq}
   \left[\frac{1}{L_\varphi^2}-\I\frac{\omega}{D}-D_x^2\right]
   P_\omega^\mathrm{(d,c)}(x,x')=\delta(x-x')
\end{equation}  
where the covariant derivative $D_x=\partial_x-2\I eA_\pm(x)$ involves the vector potential $A_\pm=\big[A\pm A']/2$ for Diffuson ($A_-$) and Cooperon ($A_+$).

\item
The penetration of the magnetic field in the wire is taken into account through the substitution $1/L_\varphi^2\to1/L_\varphi^2+1/L_{(\mathcal{B}\mp\mathcal{B}')/2}^2$, as for the WL (\S~\ref{subsec:3.1}).
  
\item  
  These expressions are based on the current conserving expressions for the conductivity correlations, given by the procedure of Kane, Serota and Lee (i.e. we combine results of Refs. \cite{AltShk86} and \cite{KanSerLee88}).

\item
  Transmission correlations are related to conductivity tensor correlations, $\sigma_{ab}=e^2\nu_0D_{ab}$. 
  The two first contributions (\ref{eq:Tucf1},\ref{eq:Tucf2}), which correlate the indices of the two tranmissions are interpreted as diffusion constant correlations $(e^2\nu_0)^2\smean{\delta D_{ab}\delta D_{cd}}$, while the two last contributions (\ref{eq:Tucf3},\ref{eq:Tucf4}), which do not correlate indices, are related to density of states fluctuations $(e^2D)^2\smean{\delta\nu^2}$~\cite{AltShk86}.

\item 
  The products of Diffusons may be related to derivative of classical transport coefficients, as it was done for the WL. For example 
  \begin{equation}
   \frac{1}{\ell_e^2}
  \derivp{ P_d(\underline{\alpha},x) }{x} 
  \derivp{P_d(\underline{\mu},x)}{x} 
  = \frac{1}{\xiloc} \derivp{ T_{\alpha\mu}^\mathrm{class}}{l_i}
  \end{equation}
  when $x$ belongs to the wire $i$,  etc.
\end{itemize}

\subsection{Four-terminal resistance correlations}

We now simplify the above expressions by neglecting the effect of thermal smearing, $\delta_T(\omega)\to\delta(\omega)$. 
Thermal effect will be described later in \S~\ref{subsec:ThermalSmearing}.
We deduce the correlations for the 4TRs (cf. \ref{app:Dem4TRF})~:
\begin{align}
\label{eq:4TRcorrel1}
  \smean{
    &\mathcal{R}_{\alpha\beta,\mu\nu}\,\mathcal{R}_{\alpha'\beta',\mu'\nu'}}^{(1)} 
    = \frac{4}{\xiloc^2}   
  \sum_{i,i'}
  \\\nonumber
  &\hspace{0.5cm} \times
  \frac{\partial\mathcal{R}^\mathrm{class}_{\mu'\nu',\mu\nu}}
       {\partial\,l_{i}}\:
  \frac{\partial\mathcal{R}^\mathrm{class}_{\alpha'\beta',\alpha\beta}}
       {\partial\,l_{i'}}
  \int_{(i)}\hspace{-0.25cm}\D x
  \int_{(i')}\hspace{-0.25cm}\D x'\:
  P_d(x,x')^2
  \\
  \label{eq:4TRcorrel2}
  \smean{
    &\mathcal{R}_{\alpha\beta,\mu\nu}\,\mathcal{R}_{\alpha'\beta',\mu'\nu'}}^{(2)} 
    = 
  \frac{4}{\xiloc^2} 
  \sum_{i,i'}
  \\\nonumber
  &\hspace{0.5cm} \times
  \frac{\partial\mathcal{R}^\mathrm{class}_{\alpha'\beta',\mu\nu}}
       {\partial\,l_{i}}\:
  \frac{\partial\mathcal{R}^\mathrm{class}_{\alpha\beta,\mu'\nu'}}
       {\partial\,l_{i'}}
  \int_{(i)}\hspace{-0.25cm}\D x
  \int_{(i')}\hspace{-0.25cm}\D x'\:
  P_c(x,x')^2
  \\
  \label{eq:4TRcorrel3}
  &\smean{
    \mathcal{R}_{\alpha\beta,\mu\nu}\,\mathcal{R}_{\alpha'\beta',\mu'\nu'}}^{(3)} 
  = 
  \frac{2}{\xiloc^2}\, 
  \sum_{i,i'}
  \\\nonumber
  &\hspace{0.5cm} \times
  \frac{\partial\mathcal{R}^\mathrm{class}_{\alpha\beta,\mu\nu}}
       {\partial\,l_{i}}\:
  \frac{\partial\mathcal{R}^\mathrm{class}_{\alpha'\beta',\mu'\nu'}}
       {\partial\,l_{i'}}
  \int_{(i)}\hspace{-0.25cm}\D x
  \int_{(i')}\hspace{-0.25cm}\D x'\:
  P_d(x,x')^2
  \\
  \label{eq:4TRcorrel4}
  &\smean{
    \mathcal{R}_{\alpha\beta,\mu\nu}\,\mathcal{R}_{\alpha'\beta',\mu'\nu'}}^{(4)} 
  = 
  \mbox{ same as }  \smean{\cdots}^{(3)} 
  \nonumber\\
  &  \hspace{4cm} \mbox{ with }  P_d\to P_c
  \:,
\end{align}
where $\xiloc=\alpha_dN_c\ell_e$ is the localisation length for the infinitely long wire.
As there will be no possible confusion, we now adopt the simpler notation $P_{d,c}=P_{\omega=0}^\mathrm{(d,c)}$.
Note that, as for the transmission correlations, only the contributions 
$\smean{\cdots}^{(1)}$ and $\smean{\cdots}^{(2)}$ correlate the indices 
in a non-trivial way.
As for the WL, the contributions of each wires are weighted by classical quantities.

\subsection{Four-terminal resistances in a multiterminal wire}

We apply our formalism to the analysis of the 4TR correlations in a wire connected to several voltage probes, like the one represented in Fig.~\ref{fig:Wire4TR}.
Let us first recall the expressions of the classical resistances, that will be
needed to compute the weights in
Eq.~(\ref{eq:4TRcorrel1},\ref{eq:4TRcorrel2},\ref{eq:4TRcorrel3},\ref{eq:4TRcorrel4})~:
\begin{align}
  \mathcal{R}^\mathrm{class}_{12,34} &= l_b/\xiloc
  \\
  \mathcal{R}^\mathrm{class}_{12,12} &= (l_a+l_b+l_c) /\xiloc
  \\
  \mathcal{R}^\mathrm{class}_{34,34} &= (l_d+l_b+l_f) /\xiloc
  \:,  
\end{align}
where $\xiloc$ is the localisation length in the infinitely long wire.
As they were considered in Ref.~\cite{BenUmbLaiWeb87}, we will study the symmetric and antisymmetic resistances $\mathcal{R}_{S,A}$, which exhibit remarkable behaviours. The relations between correlators are~:
\begin{equation}
  \smean{\delta\mathcal{R}^2_{S,A}} = 
  \frac12
  \left(
     \smean{ \delta\mathcal{R}^2_{12,34} } \pm
     \smean{ \delta\mathcal{R}_{12,34} \delta\mathcal{R}_{34,12} }
  \right)  
  \:.
\end{equation}
We will first consider the limit of strong magnetic field ($L_\mathcal{B}\ll L_\varphi$ or $l_b$), when the Cooperon contributions (\ref{eq:4TRcorrel2},\ref{eq:4TRcorrel4}) are suppressed.
The effect of a small field will be discussed in~\S~\ref{subsec:BfieldEffect}.

\subsubsection{Weakly coherent regime $L_\varphi\ll l_b$}

\paragraph{Fluctuations $\smean{ \delta\mathcal{R}^2_{12,34} }$ ---}

The contribution (\ref{eq:4TRcorrel1}) is explicitly~:
\begin{align}
    \mathcal{C}_1 &= \smean{ \delta\mathcal{R}^2_{12,34} }^{(1)}
    =\frac{4}{\xiloc^2}
    \sum_{i,j}
      \derivp{\mathcal{R}^\mathrm{class}_{12,12}}{l_i}\,
      \derivp{\mathcal{R}^\mathrm{class}_{34,34}}{l_j}\,
    \nonumber\\      
    &\hspace{1.5cm}\times\int_{(i)}\D x\int_{(j)}\D x'\,[P_d(x,x')]^2
  \:.
\end{align}
The wire weights are all equal to $1/\xiloc^2$ and imply $x\in(a),\,(b),\,(c)$ and $x'\in(d),\,(b),\,(f)$, thus this contribution can be splitted in nine terms
$\mathcal{C}_1=\mathcal{C}_{1,1}+\cdots+\mathcal{C}_{1,9}$ of four different types (\ref{eq:TermC11},\ref{eq:TermC12},\ref{eq:TermC16},\ref{eq:TermC18}).
The first term involves a double integral in the central wire
\begin{align}
    \label{eq:TermC11}
    \mathcal{C}_{1,1} 
    &\longrightarrow \diagram{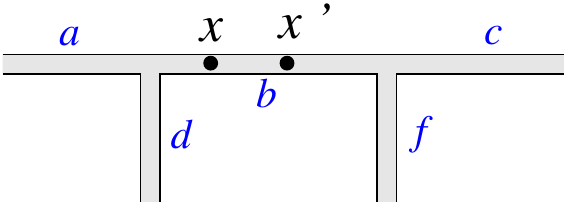}{0.75}{-0.75cm}    
    \\
    \mathcal{C}_{1,1} &= \frac{4}{\xiloc^4}
    \int_{(b)}\D x\int_{(b)}\D x'\, [P_d(x,x')]^2 
    \\\nonumber
    &\simeq 
    \frac{4}{\xiloc^4}\,l_b\int\D(x-x')\, [P_d(x,x')]^2 
    \simeq \frac{ L_\varphi^3l_b }{\xiloc^4}
    \:.   
\end{align}
where we have used the expression $P_d(x,x')\simeq(L_\varphi/2)\exp\big[-|x-x'|/L_\varphi\big]$ valid in the bulk (i.e. the expression obtained in an infinitely long wire), as the presence of the wires $(a)$, $(c)$, $(d)$ and $(f)$ only affects the diffuson at distance $\lesssim L_\varphi$ from the two vertices.

Then four terms, $\mathcal{C}_{1,2}$ to $\mathcal{C}_{1,5}$, involve integration in neighbouring wires with one coordinate in the wire $(b)$ and the other in a connecting wire, like
  \begin{equation}
    \label{eq:TermC12}
     \mathcal{C}_{1,2} \longrightarrow \diagram{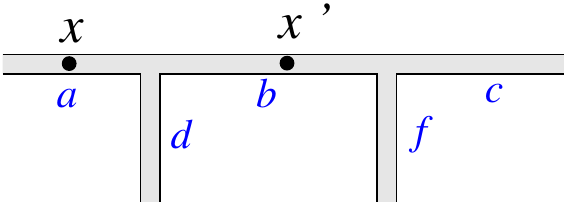}{0.75}{-0.75cm} 
     \:.
  \end{equation}
In the limit $L_\varphi\ll l_b$, integrals are dominated by $x$ and $x'$ close to the vertex and we can use the expression of the Diffuson for the star graph with coordination number $3$ given in~\ref{appendix:Star}
\begin{equation}
  \mathcal{C}_{1,2}
  \simeq \frac{4}{\xiloc^4}\int_0^\infty\D x\int_0^\infty\D x'\,
  \left(\frac{L_\varphi}{3}\EXP{-(x+x')/L_\varphi}\right)^2
\end{equation}
leading to~\footnote{
  The equality $\mathcal{C}_{1,2} =  \cdots = \mathcal{C}_{1,5}$ holds for connecting wires much longer than~$L_\varphi$.
}
\begin{equation}
    \mathcal{C}_{1,2} =  \cdots = \mathcal{C}_{1,5} \simeq 
    \frac{1}{9}\left(\frac{L_\varphi}{\xiloc}\right)^4
    \:.
\end{equation}
Two terms involve integration in neighbouring long connecting wires, like
\begin{equation}
    \label{eq:TermC16}
     \mathcal{C}_{1,6} \longrightarrow \diagram{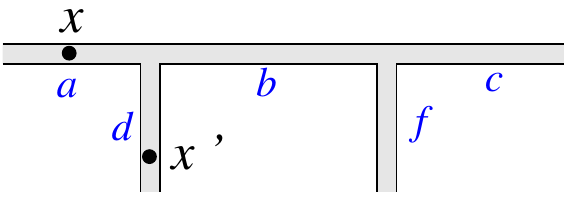}{0.75}{-0.75cm} 
     \:.
\end{equation}
When $L_\varphi\ll l_b$, we have $\mathcal{C}_{1,6}= \mathcal{C}_{1,7}\simeq\mathcal{C}_{1,2}$.
The two last contributions are of the kind
\begin{equation}
    \label{eq:TermC18}
     \mathcal{C}_{1,8} \longrightarrow \diagram{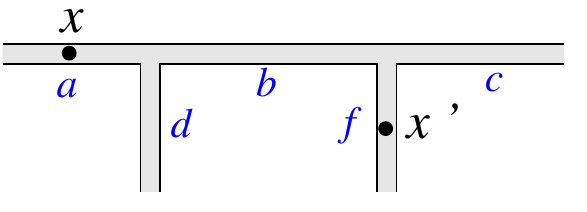}{0.75}{-0.75cm}
\end{equation}
and are exponentially suppressed
\begin{equation}
    \mathcal{C}_{1,8} = \mathcal{C}_{1,9} \sim  
     \left(\frac{L_\varphi}{\xiloc}\right)^4 \EXP{-2l_b/L_\varphi}
    \:.
\end{equation}

The contribution  $\smean{ \delta\mathcal{R}^2_{12,34} }^{(3)}$ is simpler to discuss as it involves the weight
\begin{equation}
    \derivp{\mathcal{R}^\mathrm{class}_{12,34}}{l_i}\,
    \derivp{\mathcal{R}^\mathrm{class}_{12,34}}{l_j}
\end{equation}
leading to the constraint $x,\:x'\in(b)$.
Thus we immediately get
\begin{equation}
  \smean{ \delta\mathcal{R}^2_{12,34} }^{(3)} = \frac12 \mathcal{C}_{1,1}
  \:.
\end{equation}

Summing the ten terms, we get the fluctuations 
$
\smean{ \delta\mathcal{R}^2_{12,34} }
=(3/2)\,\mathcal{C}_{1,1}
+\mathcal{C}_{1,2}+\cdots+\mathcal{C}_{1,9}
$.
Since the first term dominates, we recover the well known expression~\cite{AkkMon07} for the conductance fluctuations 
\begin{equation}
  \label{eq:UCFwellknown1}
  \frac{ \smean{ \delta\mathcal{R}^2_{12,34} } }{ (\mathcal{R}^\mathrm{class}_{12,34})^4}
  \simeq 
  \frac32\,\left( \frac{L_\varphi}{l_b} \right)^3 \ll 1
  \:.
\end{equation}

\paragraph{Correlations $\smean{ \delta\mathcal{R}_{12,34} \delta\mathcal{R}_{34,12} }$ ---}

The weights of the contribution $\smean{ \cdots }^{(1)}$ are 
\begin{equation}
    \label{eq347}
    \derivp{\mathcal{R}^\mathrm{class}_{12,34}}{l_i}\,
    \derivp{\mathcal{R}^\mathrm{class}_{34,12}}{l_j}\,
\end{equation}
which requires once again $x,\,x'\in(b)$. Therefore 
$
    \smean{ \delta\mathcal{R}_{12,34}\delta\mathcal{R}_{34,12} }^{(1)}
    =\mathcal{C}_{1,1}
$.

The contribution $\smean{ \cdots }^{(3)}$ takes the same form, hence
$
    \smean{ \delta\mathcal{R}_{12,34}\delta\mathcal{R}_{34,12} }^{(3)}
    =(1/2)\mathcal{C}_{1,1}
$.

The correlations are given by summing these two terms
$\smean{ \delta\mathcal{R}_{12,34}\delta\mathcal{R}_{34,12} } = (3/2)\, \mathcal{C}_{1,1}$, which coincides exactly with the dominant term of the fluctuations.

\paragraph{Conclusion ---}
If we now gather all these results we obtain
\begin{align}
  \label{eq:DecompositionRS}
  \smean{\delta\mathcal{R}^2_{S}} 
  &=  \frac{
  3\,\mathcal{C}_{1,1}+\mathcal{C}_{1,2}+\cdots+\mathcal{C}_{1,9} }{2}
  \simeq \frac32\frac{L_\varphi^3l_b}{\xiloc^4} 
  \\
  \label{eq:DecompositionRA}
  \smean{\delta\mathcal{R}^2_{A}} 
  &=
  \frac{\mathcal{C}_{1,2}+\cdots+\mathcal{C}_{1,9} }{2}
  \simeq  \frac13 \left(\frac{L_\varphi}{\xiloc}\right)^4
  \:.
\end{align}
$\delta\mathcal{R}_{S}$ grows with $l_b$ whereas $\delta\mathcal{R}_{A}$ is independent on the distance between the two voltage probes, in agreement with the experiment (see figure~\ref{fig:benoit}).

\subsubsection{Coherent limit $l_b\ll L_\varphi$}

We now discuss the case where the central wire is coherent, with long connecting wires ($l_a,\,l_c,\,l_d,\,l_f\gg L_\varphi$). We can write that the Diffuson is almost uniform inside the wire $(b)$, equal to the Diffuson at a vertex of coordination number $4$. Thus $P_d(x,x')\simeq L_\varphi/4$ for $x,\,x'\in(b)$.
It decays exponentially over the distance $L_\varphi$ in the connecting wires~:
$P_d(x,x')\simeq(L_\varphi/4)\,\exp[-(x+x')/L_\varphi]$ for $x\in(a)$ and $x'\in(c)$, etc. (cf.~\ref{appendix:Star}).
As a result we see that the fluctuations are dominated by four terms, $\mathcal{C}_{1,6},\cdots,\mathcal{C}_{1,9}$ corresponding to cases like \eqref{eq:TermC16} or \eqref{eq:TermC18}, when $x$ and $x'$ are both integrated over the long distance $L_\varphi$ in the connecting wires~:
$\smean{ \delta\mathcal{R}_{12,34}^2 }\simeq(1/4)\,(L_\varphi/\xiloc)^4$, i.e. the four-terminal conductance fluctuations are large
\begin{equation}
  \label{eq:UCFlarge}
  \frac{\smean{ \delta\mathcal{R}^2_{12,34} }}{(\mathcal{R}^\mathrm{class}_{12,34})^4}
  \simeq\frac14\left(\frac{L_\varphi}{l_b}\right)^4\gg1
\:.
\end{equation}
The correlations are much smaller as they involve integrations of $x$ and $x'$ in the central wire only
$\smean{ \delta\mathcal{R}_{12,34} \delta\mathcal{R}_{34,12} }\simeq(3/8)\,L_\varphi^2l_b^2/\xiloc^4$, i.e.
\begin{equation}
  \frac{\smean{ \delta\mathcal{R}_{12,34} \delta\mathcal{R}_{34,12} }}{(\mathcal{R}^\mathrm{class}_{12,34}\mathcal{R}^\mathrm{class}_{34,12})^2}
  \simeq\frac38\left(\frac{L_\varphi}{l_b}\right)^2\gg1
  \:.
\end{equation}
Finally we get
\begin{equation}
  \label{eq:FluctSAcoherent}
  \smean{\delta\mathcal{R}^2_{S}} \simeq \smean{\delta\mathcal{R}^2_{A}}
  \simeq 
  \frac{1}{8}\left(\frac{L_\varphi}{\xiloc}\right)^4
  \:.
\end{equation}
Fluctuations are independent on the distance $l_b$ between the voltage probes.

\subsubsection{Thermal smearing}
\label{subsec:ThermalSmearing}

In the two previous paragraphs, we have reproduced the main conclusions of Hershfield \cite{Her89} by simpler arguments based on the analysis of the wire weights.
Although the qualitative change of behaviour between $\delta\mathcal{R}_{S}$ and $\delta\mathcal{R}_{A}$ at $l_b\sim L_\varphi$ agrees with the experiment of Benoit \textit{et al.}~\cite{BenUmbLaiWeb87} (Fig.~\ref{fig:benoit}), it was noticed that thermal smearing, not accounted for by Hershfield, is important in the experiment. This corresponds to the case where the thermal length $L_T=\sqrt{D/T}$ is smaller than $L_\varphi$.

There is also a fundamental reason to consider this regime~: in the low temperature regime ($T\lesssim1\:\mathrm{K}$) and in the absence of external degrees of freedom like magnetic impurities, the decoherence is dominated by electronic interactions. As a consequence one has $L_T\ll L_\varphi$ (see~\cite{AltAro85,TexMon05b,AkkMon07,TexDelMon09,Tex10hdr} and references therein).

In this regime the contribution $\mean{\cdots}^{(3)}$ is negligible compared to  $\mean{\cdots}^{(1)}$.

\paragraph{Weakly coherent limit $L_T\ll L_\varphi\ll l_b$ ---}

We should repeat the analysis of the previous subsection by adding an integration over the frequency with the thermal function.
The first contribution to the resistance fluctuations reads
\begin{align}
    \mathcal{C}_{1,1} 
    &= \frac{4}{\xiloc^4}
    \int\D\omega\,\delta_T(\omega)
    \\\nonumber
    &\times\int_{(b)}\D x\int_{(b)}\D x'\,  
    P_\omega^\mathrm{(d)}(x,x')P_{-\omega}^\mathrm{(d)}(x,x')
    \:.
\end{align}
At finite temperature, the Diffuson involves exponentials like 
$\exp[-\sqrt{\tilde{\gamma}}x]$ with $\tilde{\gamma}=1/L_\varphi^2-\I\omega/D$ where $\omega/D\lesssim1/L_T^2$.
When $\mathrm{min}(L_T,L_\varphi)\ll l_b$, the Diffuson decays rapidly inside the wire.
$\mathcal{C}_{1,1}$
is dominated by the integral in the bulk of the wire, which leads to a similar calculation as for an infinite wire.
When $L_T\ll L_\varphi$  (the case $L_\varphi\ll L_T\ll l_b$ is similar to $L_T=\infty$, and was discussed above), the thermal function can be considered as a broad function and simply replaced by $\delta_T(\omega)\to1/(6T)$.
Using the expression of the Diffuson in bulk we get
\begin{equation}
  \mathcal{C}_{1,1} \simeq \frac{\pi}{3}\,\frac{L_T^2L_\varphi l_b}{\xiloc^4}
  \:.
\end{equation}
This term dominates the fluctuations, therefore we recover the known expression~\cite{AkkMon07} of the conductance fluctuations
\begin{equation}
  \label{eq:UCFwellknown2}
  \frac{ \smean{\delta\mathcal{R}^2_{12,34}} }{ (\mathcal{R}^\mathrm{class}_{12,34})^4 }
  \simeq \frac{\pi}{3}\, \frac{ L_T^2L_\varphi }{ l_b^3 }
  \:.
\end{equation}

Next contributions are the six terms of type \eqref{eq:TermC12} and \eqref{eq:TermC16}. We can again argue that integration is dominated by $x$ and $x'$ close to the vertex and use the approximate form given in \ref{appendix:Star}~:
\begin{align}
    \mathcal{C}_{1,2} 
    &\simeq \frac{4}{\xiloc^4}
    \int\frac{\D\omega}{6T}
    \\\nonumber
    &\times\int_{(a)}\D x\int_{(b)}\D x'\,  
    P_\omega^\mathrm{(d)}(x,x')P_{-\omega}^\mathrm{(d)}(x,x')
    \:.   
\end{align}
where $P_\omega^\mathrm{(d)}(x,x')\simeq1/(3\sqrt{\tilde{\gamma}})\EXP{-\sqrt{\tilde{\gamma}}(x+x')}$ with $\tilde{\gamma}=1/L_\varphi^2-\I\omega/D$.
Some algebra gives
\begin{equation}
  \label{eq:TermC12WithThermalSmearing}
  \mathcal{C}_{1,2} \simeq \cdots \simeq \mathcal{C}_{1,7} 
  \simeq \frac{2}{27}\,\frac{L_T^2L_\varphi^2}{\xiloc^4}
  \:.
\end{equation}

Correlations are obtained by similar arguments
$\smean{ \delta\mathcal{R}_{12,34}\delta\mathcal{R}_{34,12} }=\mathcal{C}_{1,1}$, which coincides with the dominant term of the fluctuations.

Going back to the quantity of interest, we should replace \eqref{eq:DecompositionRS} by 
$\smean{\delta\mathcal{R}^2_{S}} = \mathcal{C}_{1,1}+(\mathcal{C}_{1,2}+\cdots+\mathcal{C}_{1,9})/2$ whereas \eqref{eq:DecompositionRA} still holds, thus
\begin{align}
  \smean{\delta\mathcal{R}^2_{S}} & \simeq \frac{\pi}{3}\frac{L_T^2L_\varphi l_b}{\xiloc^4}
  \\
  \smean{\delta\mathcal{R}^2_{A}} & \simeq \frac{2}{9}\frac{L_T^2L_\varphi^2}{\xiloc^4}
  \:.
\end{align}
We reproduce similar conclusions as for $L_T=\infty$, i.e. fluctuations $\delta\mathcal{R}_S$ growing with the distance $l_b$ between the voltage probes, while the fluctuations  $\delta\mathcal{R}_A$ are independent on~$l_b$.

\paragraph{Coherent limit $l_b\ll L_T\ll L_\varphi$ ---}

As in the case $l_b\ll L_\varphi\ll L_T$ analyzed previously, the fluctuations are dominated by four terms where $x$ and $x'$ are integrated over two long connecting wires. 
It is simply given by multiplying \eqref{eq:TermC12WithThermalSmearing} by a factor $(3/4)^2$, which accounts for the fact that the Diffuson feels an effective coordination number $4$ instead of $3$, hence $\mathcal{C}_{1,6}\simeq\cdots\simeq\mathcal{C}_{1,9}\simeq(1/24)L_T^2L_\varphi^2/\xiloc^4$.
Thus
\begin{equation}
  \label{eq:UCFlargeWithThermalBroadening}
  \frac{\smean{ \delta\mathcal{R}^2_{12,34} }}{(\mathcal{R}^\mathrm{class}_{12,34})^4}
  \simeq\frac{1}{6}\left(\frac{L_T}{l_b}\right)^2\left(\frac{L_\varphi}{l_b}\right)^2
  \gg1
  \:.
\end{equation}
The analysis of the correlations 
$\smean{ \delta\mathcal{R}_{12,34}\delta\mathcal{R}_{34,12} }=\mathcal{C}_{1,1}$
is as follows~:
we write $P^\mathrm{(d)}_\omega(x,x')\simeq1/(4\sqrt{\tilde{\gamma}})$ when $x,\,x'\in(b)$, therefore
$\mathcal{C}_{1,1}\simeq(1/4)(l_b^2/\xiloc^4)\int\D\omega\,\delta_T(\omega)/|\tilde{\gamma}|$ where $\tilde{\gamma}=1/L_\varphi^2-\I\omega/D$.
The presence of the function $\delta_T(\omega)$ is needed in order to cut off the contribution of large frequencies. When $L_T\ll L_\varphi$, some algebra gives $\mathcal{C}_{1,1}\simeq(1/6)(L_T^2l_b^2/\xiloc^4)\ln(L_\varphi/L_T)$, leading to 
\begin{equation}
  \frac{\smean{ \delta\mathcal{R}_{12,34} \delta\mathcal{R}_{34,12} }}{(\mathcal{R}^\mathrm{class}_{12,34}\mathcal{R}^\mathrm{class}_{34,12})^2}
  \simeq\frac16\left(\frac{L_T}{l_b}\right)^2\ln(L_\varphi/L_T)
  \:.
\end{equation}
Quite surprisingly, we obtain a logarithmic dependence in $L_\varphi$ reminiscent of the 2D situation, although the system is 1D.

The symmetric and antisymmetric resistance fluctuations are
\begin{equation}
  \smean{\delta\mathcal{R}^2_{S}} \simeq  \smean{\delta\mathcal{R}^2_{A}}  
  \simeq \frac{1}{12}\frac{L_T^2L_\varphi^2}{\xiloc^4}
  \:.
\end{equation}

\subsubsection{Magnetic field dependence}
\label{subsec:BfieldEffect}

Finally, we discuss the magnetic field dependence.
The above results are valid when the magnetic field $\mathcal{B}=\mathcal{B}'$ is larger than the correlation field, i.e. in a diffusive wire, when $\mathcal{B}\gg\mathcal{B}_c\sim\phi_0/(L_\varphi w)$ where $w$ is the width of the wires and $\phi_0=h/e$ the flux quantum.
At small magnetic field, Cooperon contributions (\ref{eq:4TRcorrel2},\ref{eq:4TRcorrel4}) must be taken into account as well.
This can be done easily by noticing that, due to the wire weights, the role of fluctuations and correlations are exchanged for the two first contributions~:
\begin{align}
 \nonumber
   \smean{ \delta\mathcal{R}^2_{12,34} }^{(2)}
   &= \smean{ \delta\mathcal{R}_{12,34}\delta\mathcal{R}_{34,12} }^{(1)}\big|_{P_d\to P_c}
   \\
 \nonumber
   \smean{ \delta\mathcal{R}_{12,34}\delta\mathcal{R}_{34,12} }^{(2)}
   &= \smean{ \delta\mathcal{R}^2_{12,34} }^{(1)}\big|_{P_d\to P_c}
\end{align}
while 
\begin{align}
 \nonumber
   \smean{ \delta\mathcal{R}^2_{12,34} }^{(4)}
   &= \smean{ \delta\mathcal{R}^2_{12,34} }^{(3)}\big|_{P_d\to P_c}
   \\
 \nonumber
   \smean{ \delta\mathcal{R}_{12,34}\delta\mathcal{R}_{34,12} }^{(4)}
  &= 
  \smean{ \delta\mathcal{R}_{12,34}\delta\mathcal{R}_{34,12} }^{(3)}\big|_{P_d\to P_c}
  \:.
\end{align}
The substitution $P_d\to P_c$ is simply achieved by doing $1/L_\varphi^2\to1/L_\varphi^2+1/L_\mathcal{B}^2$, cf. \S~\ref{subsec:3.1} (the analysis of the correlations for different fields requires similar substitution in both the Diffuson and the Cooperon, cf. \S~\ref{subsec:4.1}).
We deduce straightforwardly the crossover functions.
We only consider the weakly coherent regime, $L_\varphi\ll l_b$.

\paragraph{Regime $L_\varphi\ll l_b,\ L_T$ ---}

Using the previous calculations we obtain 
\begin{align}
  \smean{\delta\mathcal{R}^2_{S}}
  &\simeq 
   \frac32 \frac{L_\varphi^3l_b}{\xiloc^4} \left(
      1+ \left[ 1 + (\mathcal{B}/\mathcal{B}_c)^2 \right]^{-3/2}
   \right)
  \\
  \smean{\delta\mathcal{R}^2_{A}} 
  &\simeq \frac13\left(\frac{L_\varphi}{\xiloc}\right)^4
  \left(  1 - \left[ 1 + (\mathcal{B}/\mathcal{B}_c)^2 \right]^{-2} \right)
  \:,
\end{align}
where $\mathcal{B}/\mathcal{B}_c=L_\varphi/L_\mathcal{B}=(2\pi/\sqrt{3})\mathcal{B}wL_\varphi/\phi_0$.
Whereas the symmetric resistance fluctuations are doubled at small field,
the antisymmetric resistance vanishes as 
$\delta\mathcal{R}_{A}(\mathcal{B})\sim(L_\varphi/\xiloc)^2\,(\mathcal{B}L_\varphi w/\phi_0)$ for $\mathcal{B}\to0$.

\paragraph{Regime $L_T\ll L_\varphi\ll l_b$ ---}

In this case we obtain 
\begin{align}
  \smean{\delta\mathcal{R}^2_{S}}
  &\simeq 
   \frac{\pi}{3} \frac{L_T^2 L_\varphi l_b}{\xiloc^4} \left(
      1+ \left[ 1 + (\mathcal{B}/\mathcal{B}_c)^2 \right]^{-1/2}
   \right)
  \\
  \smean{\delta\mathcal{R}^2_{A}} 
  &\simeq \frac1{12}\frac{L_T^2 L_\varphi^2}{\xiloc^4}
  \frac{\mathcal{B}^2}{\mathcal{B}_c^2+\mathcal{B}^2}
  \:.
\end{align}
The antisymmetric resistance now vanishes at small field as 
$\delta\mathcal{R}_{A}(\mathcal{B})\sim(L_TL_\varphi/\xiloc^2)\,(\mathcal{B}L_\varphi w/\phi_0)$ for $\mathcal{B}\to0$.

\subsubsection{Experiments}

\begin{figure}[!ht]
\centering
\begin{picture}(0,0)%
\includegraphics{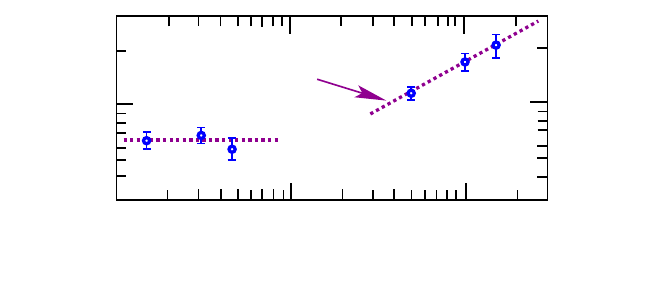}%
\end{picture}%
\setlength{\unitlength}{4144sp}%
\begingroup\makeatletter\ifx\SetFigFont\undefined%
\gdef\SetFigFont#1#2#3#4#5{%
  \reset@font\fontsize{#1}{#2pt}%
  \fontfamily{#3}\fontseries{#4}\fontshape{#5}%
  \selectfont}%
\fi\endgroup%
\begin{picture}(3026,1370)(6425,-4761)
\put(7359,-3782){\makebox(0,0)[lb]{\smash{{\SetFigFont{12}{14.4}{\rmdefault}{\mddefault}{\updefault}{\color[rgb]{.56,0,.56}$\propto\sqrt{L}$}%
}}}}
\put(6696,-3904){\makebox(0,0)[lb]{\smash{{\SetFigFont{10}{12.0}{\rmdefault}{\mddefault}{\updefault}{\color[rgb]{0,0,0}100}%
}}}}
\put(6771,-4330){\makebox(0,0)[lb]{\smash{{\SetFigFont{10}{12.0}{\rmdefault}{\mddefault}{\updefault}{\color[rgb]{0,0,0}30}%
}}}}
\put(7593,-4683){\makebox(0,0)[lb]{\smash{{\SetFigFont{12}{14.4}{\rmdefault}{\mddefault}{\updefault}{\color[rgb]{0,0,0}$L$ ($\mu\mathrm{m}$)}%
}}}}
\put(8464,-4475){\makebox(0,0)[lb]{\smash{{\SetFigFont{10}{12.0}{\rmdefault}{\mddefault}{\updefault}{\color[rgb]{0,0,0}10}%
}}}}
\put(7709,-4480){\makebox(0,0)[lb]{\smash{{\SetFigFont{10}{12.0}{\rmdefault}{\mddefault}{\updefault}{\color[rgb]{0,0,0}1}%
}}}}
\put(6867,-4477){\makebox(0,0)[lb]{\smash{{\SetFigFont{10}{12.0}{\rmdefault}{\mddefault}{\updefault}{\color[rgb]{0,0,0}0.1}%
}}}}
\put(6696,-3512){\makebox(0,0)[lb]{\smash{{\SetFigFont{10}{12.0}{\rmdefault}{\mddefault}{\updefault}{\color[rgb]{0,0,0}300}%
}}}}
\put(6608,-4334){\rotatebox{90.0}{\makebox(0,0)[lb]{\smash{{\SetFigFont{12}{14.4}{\rmdefault}{\mddefault}{\updefault}{\color[rgb]{0,0,0}$\delta\mathcal{R}_{12,34}$ ($\Omega$)}%
}}}}}
\end{picture}%
\caption{\it Mesoscopic (sample to sample) resistance fluctuations for multiconnected silicon inversion-layer narrow wire at $T\simeq400\:\mathrm{mK}$.
The two samples have lengths $L\simeq0.15\:\mu\mathrm{m}$ ($\mathcal{R}_{12,34}\approx450\:\Omega$) and $5\:\mu\mathrm{m}$ ($\mathcal{R}_{12,34}\approx12\:\mathrm{k}\Omega$). 
Data from Ref.~\cite{SkoManHowJacTenSto87}.}
\label{fig:skocpol}
\end{figure}

We now rediscuss the experiments at the light of our results.
In Fig.~\ref{fig:skocpol}, we have reproduced the experimental data obtained by Skocpol \textit{et al.}~\cite{SkoManHowJacTenSto87} for a Silicon inversion-layer narrow wire.
We now denote by $L$ the distance between the two voltage probes (denoted $l_b$ above).
The experimental result exhibits the behaviours obtained above~: 
a growth of the fluctuations with the length, $\delta\mathcal{R}_{12,34}\sim L_\varphi^{3/2}L^{1/2}/\xiloc^2$ in the incoherent regime $L_\varphi\lesssim L$, Eq.~\eqref{eq:UCFwellknown1}, and a saturation $\delta\mathcal{R}_{12,34}\sim(L_\varphi/\xiloc)^{2}$ in the coherent regime $L_\varphi\gtrsim L$, Eq.~\eqref{eq:UCFlarge}.
Note that accounting for thermal broadening does not change this conclusion, cf. Eqs.~\eqref{eq:UCFwellknown2} and \eqref{eq:UCFlargeWithThermalBroadening}.

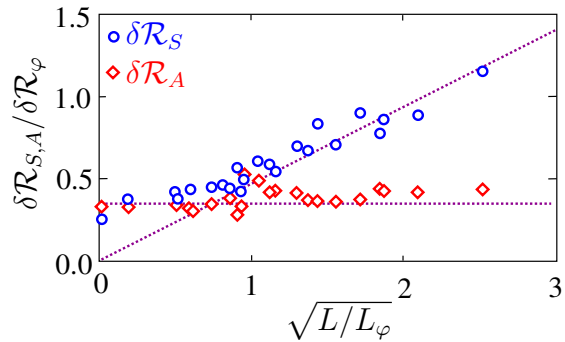
\begin{figure}[!ht]
\centering
\begin{picture}(0,0)%
\includegraphics{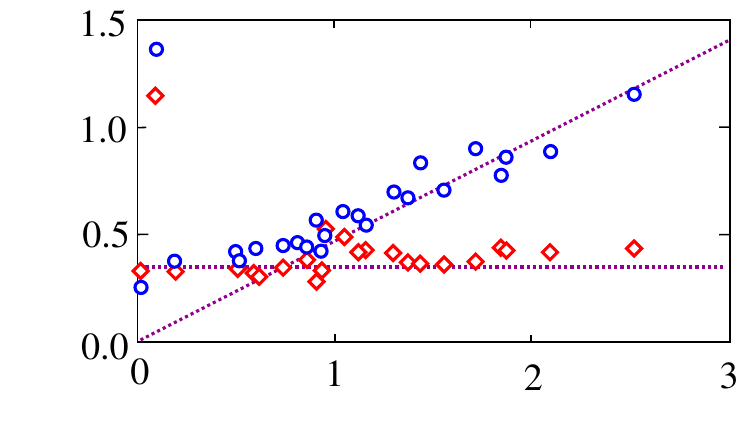}%
\end{picture}%
\setlength{\unitlength}{3947sp}%
\begingroup\makeatletter\ifx\SetFigFont\undefined%
\gdef\SetFigFont#1#2#3#4#5{%
  \reset@font\fontsize{#1}{#2pt}%
  \fontfamily{#3}\fontseries{#4}\fontshape{#5}%
  \selectfont}%
\fi\endgroup%
\begin{picture}(3600,2138)(4201,-5475)
\put(6018,-5397){\makebox(0,0)[lb]{\smash{{\SetFigFont{12}{14.4}{\rmdefault}{\mddefault}{\updefault}{\color[rgb]{0,0,0}$\sqrt{L/L_\varphi}$}%
}}}}
\put(4471,-4646){\rotatebox{90.0}{\makebox(0,0)[lb]{\smash{{\SetFigFont{12}{14.4}{\rmdefault}{\mddefault}{\updefault}{\color[rgb]{0,0,0}$\delta\mathcal{R}_{S,A}/\delta\mathcal{R}_\varphi$}%
}}}}}
\put(5027,-3848){\makebox(0,0)[lb]{\smash{{\SetFigFont{12}{14.4}{\rmdefault}{\mddefault}{\updefault}{\color[rgb]{1,0,0}$\delta\mathcal{R}_{A}$}%
}}}}
\put(5026,-3608){\makebox(0,0)[lb]{\smash{{\SetFigFont{12}{14.4}{\rmdefault}{\mddefault}{\updefault}{\color[rgb]{0,0,1}$\delta\mathcal{R}_{S}$}%
}}}}
\end{picture}%
  \caption{\it 
           Mesoscopic resistance fluctuations for multiconnected Au and Sb wires at $T=40\:\mathrm{mK}$ and $300\:\mathrm{mK}$.
           Length $L$ varies from $0.2\:\mu\mathrm{m}$ to $4\:\mu\mathrm{m}$.
           $L$ is the distance between the voltage probes.
           Blue circles correspond to the symmetric resistance
           $\smean{\delta\mathcal{R}^2_{S}}^{1/2}$ and red diamonds to the
           antisymmetric resistance
           $\smean{\delta\mathcal{R}^2_{A}}^{1/2}$. 
           The inset shows the Sb sample~;
           $\delta\mathcal{R}_\varphi=(L_\varphi/\xiloc)^2$ are the fluctuations for a wire of length $L_\varphi$.
           Data from Ref.~\cite{BenUmbLaiWeb87}.}
  \label{fig:benoit}
\end{figure}

Another remarkable experiment is the one of Benoit \textit{et al.}~\cite{BenUmbLaiWeb87}, who analyzed voltage fluctuations in Sb and Au narrow wires.
Using the data given in this reference, we have plotted the fluctuations of the symmetric and antisymmetric resistances in Fig.~\ref{fig:benoit}.
The most striking outcome is the comparison of the different behaviours for the symmetric and antisymmetric resistances.

Let us analyze the behaviours more into details. 
We follow the line of Ref.~\cite{BenUmbLaiWeb87}, where thermal broadening was not taken into account.
Voltage (sample to sample) fluctuations were measured respectively to the fluctuations for a wire of length $L_\varphi$, which coincides with the relative resistance fluctuations $\delta\mathcal{R}_{S,A}/\delta\mathcal{R}_\varphi$ where $\delta\mathcal{R}_\varphi=(L_\varphi/\xiloc)^2$.

In the coherent limit ($L_\varphi\gtrsim L$), Eq.~\eqref{eq:FluctSAcoherent} gives the value
$\delta\mathcal{R}_{S,A}/\delta\mathcal{R}_\varphi\simeq1/\sqrt{8}\simeq0.35$ consistent with the experimental data.

In the incoherent limit ($L_\varphi\lesssim L$), Eq.~\eqref{eq:DecompositionRA} gives 
$\delta\mathcal{R}_{A}/\delta\mathcal{R}_\varphi\simeq1/\sqrt{3}\simeq0.577$, whereas \eqref{eq:DecompositionRS} leads to 
$\delta\mathcal{R}_{S}/\delta\mathcal{R}_\varphi\simeq\sqrt{3/2}\,\sqrt{L/L_\varphi}$.
Although the linear behaviour with $\sqrt{L/L_\varphi}$ agrees qualitatively with the experimental data, the prefactor obtained experimentally, $\approx0.47$, is significantly smaller than $\sqrt{3/2}\simeq1.22$.

The fact that the experimental values are smaller than the theoretical predictions could be explained by the effect of thermal broadening, as it brings a reduction factor~$L_T/L_\varphi$.
A more precise analysis would nevertheless be needed. 
Another aspect which could explain this quantitative disagreement concerns the determination of the phase coherence length, which was not obtained by a unique independent procedure in Ref.~\cite{BenUmbLaiWeb87}. 
At the highest temperatures, the phase coherence length was obtained by independent WL measurements, which does not account for the fact that, in heavy metals like gold, when spin-orbit scattering is strong and magnetic impurities present, the phase coherence length involved in the weak localisation \cite{HikLarNag80} and the conductance fluctuations \cite{ChaSanPro90} differ (see also \cite{AkkMon07}).
The lack of a formula for the WL correction in the multiterminal wire did not allow the authors to extract $L_\varphi$ for the lowest temperatures.
A more reliable procedure would have been to obtain $L_\varphi$ from another long wire ($\gg L_\varphi$) made under the same conditions.


\section{Conclusion}
\label{sec:Conclu}

Since the pioneering work of Markus B\"uttiker, four-terminal resistance is now recognized as a quantity of major importance in mesoscopic physics (Ref.~\cite{But86a} is indeed B\"uttiker's most cited article, with now almost 2000 citations).
After having briefly reviewed several aspects of four-terminal resistances, we have focused the discussion on the analysis of quantum transport in networks of weakly disordered wires, for which explicit expressions for the quantum corrections to the conductances and the four-terminal resistances were discussed.
We have seen that both the weak localisation correction and the correlations involve integrals of the Cooperon (and also of the Diffuson for the correlations) inside the wires, whose contributions must be weighted by derivatives of the classical coefficients~: $\partial{G}_{\alpha\beta}^\mathrm{class}/\partial l_i$ for the conductances and $\partial\mathcal{R}_{\alpha\beta,\mu\nu}^\mathrm{class}/\partial l_i$ for the four-terminal resistances. 
Although we had discussed this earlier for the WL correction to the conductance \cite{TexMon04}, we have provided here a new interpretation of these coefficients in terms of ``\textit{generalised conductances}'' $\mathcal{G}_{i,\alpha}$ relating the current in a wire $i$ \textit{inside} the network to the external potential $V_\alpha$, precisely~:
$\partial{G}_{\alpha\beta}^\mathrm{class}/\partial l_i\propto\mathcal{G}_{i,\alpha}\mathcal{G}_{i,\beta}$.

We have illustrated the efficiency of our formalism by considering simple examples.
Our determination of the four-terminal resistance correlations only involves simple calculations (which greatly simplify the analysis of Ref.~\cite{Her89} in particular).
Moreover we have been able to consider the effect of thermal broadening, which was not studied so far.
All the main dependences are summarized in Tab.~\ref{tab:Summary1} and Tab.~\ref{tab:Summary2}.

\begin{table}[!ht]
\hspace{-0.5cm}
\begin{tabular}{|c|c|cc|}
\hline
 & $L_\varphi\ll l_b$ & \multicolumn{2}{c|}{$l_b\ll L_\varphi$} \\
 \hline
 & fluct. $\simeq$ correl.
 & fluct. & $\gg$  correl. \\
 \hline
 &&&\\[-0.25cm]
$L_\varphi\ll L_T$ & $(L_\varphi/l_b)^3$ &  $(L_\varphi/l_b)^4$ 
                   & $(L_\varphi/l_b)^2$ \\[0.2cm]
$L_T\ll L_\varphi$ & $\big(\frac{L_T}{l_b}\big)^2\frac{L_\varphi}{l_b}$ 
                   & $\big(\frac{L_T}{l_b}\big)^2\big(\frac{L_\varphi}{l_b}\big)^2$ 
                   & $\big(\frac{L_T}{l_b}\big)^2\,\ln\frac{L_\varphi}{L_T}$ \\[0.2cm] 
\hline
\end{tabular}
\caption{\it Main behaviours of the four-terminal conductance fluctuations $\smean{\delta\mathcal{R}_{12,34}^2}/(\mathcal{R}_{12,34}^\mathrm{class})^4$ and the conductance correlations $\smean{\delta\mathcal{R}_{12,34}\delta\mathcal{R}_{34,12}}/(\mathcal{R}_{12,34}^\mathrm{class})^4$ of the four-terminal resistances.
Precise numerical prefactors are given in the text.}
\label{tab:Summary1}
\end{table}

\begin{table}[!ht]
\hspace{-0.5cm}
\begin{tabular}{|c|cc|c|}
\hline
 & \multicolumn{2}{c|}{$L_\varphi\ll l_b$} & $l_b\ll L_\varphi$ \\
 \hline
 & $\smean{\delta\mathcal{R}^2_{S}}$ &$\gg$  $\smean{\delta\mathcal{R}^2_{A}}$ 
 &  $\smean{\delta\mathcal{R}^2_{S}}\simeq\smean{\delta\mathcal{R}^2_{A}}$ \\
 \hline
 &&&\\[-0.25cm]
$L_\varphi\ll L_T$ & $\big(\frac{L_\varphi}{\xiloc}\big)^3\frac{l_b}{\xiloc}$ 
                   & $(L_\varphi/\xiloc)^4$ 
                   & $(L_\varphi/\xiloc)^4$ \\[0.2cm]
$L_T\ll L_\varphi$ & $\big(\frac{L_T}{\xiloc}\big)^2\frac{L_\varphi}{\xiloc}\frac{l_b}{\xiloc}$ 
                   & $\big(\frac{L_T}{\xiloc}\big)^2\big(\frac{L_\varphi}{\xiloc}\big)^2$ 
                   & $\big(\frac{L_T}{\xiloc}\big)^2\big(\frac{L_\varphi}{\xiloc}\big)^2$  \\ [0.2cm]
\hline
\end{tabular}
\caption{\it Main behaviours of the symmetric and antisymmetric part of the four-terminal resistances. $\xiloc=\alpha_d N_c\ell_e$ is the localisation length of the wire (we recall that the validity of the diagrammatic appraoch is $\xiloc\gg\min(l_b,L_\varphi)$.
Precise numerical prefactors are given in the text.}
\label{tab:Summary2}
\end{table}



\begin{appendix}

\section{Conductance matrix, resistance matrix and four-terminal resistances}
\label{app:CRFT}

\subsection{Conductance and resistance matrices}

A natural way to characterize the linear response of a multiterminal structure is to introduce the conductance matrix relating the voltages at the contacts to the currents~:
$I_\alpha=\sum_\beta G_{\alpha\beta}V_\beta$.
The matrix elements of the conductance matrix are not independent as they must satisfy two types of constraints~: (\textit{i}) current is conserved, thus $\sum_\alpha I_\alpha=0$ whatever the choice of external potentials. (\textit{ii}) A global shift of all voltages $V_\alpha\to V_\alpha+U_0$ does not change the current.
As a consequence $\sum_\alpha G_{\alpha\beta}=\sum_\beta G_{\alpha\beta}=0$.
In other terms, introducing the vector $X_0^\mathrm{T}=(1,1,\cdots,1)$ where $^\mathrm{T}$ denotes transposition, we rewrite the second condition as $GX_0=0$, i.e. $G$ is a linear map acting in the vector space $\mathcal{E}_\perp=\{X\in\mathbb{R}^N\:|\:X_0^\mathrm{T}\cdot X=0\}$. The current conservation rewrites $X_0^\mathrm{T}G=0$, meaning that $G$ maps $\mathcal{E}_\perp$ onto itself.
In the subspace $\mathcal{E}_\perp$, we may invert the conductance matrix, which leads to introduce the resistance matrix $R$ relating the external currents to the voltages $V=RI$.
The relation between the two matrices is thus $RG=GR=\mathbf{1}_N-P_\parallel$, where $P_\parallel=(1/N)X_0X_0^\mathrm{T}$ is the projector on the vector $X_0$, where $N$ is the number of contacts. 
We write more conveniently~:
\begin{align}
  \sum_\gamma R_{\alpha\gamma} G_{\gamma\lambda} 
  = \sum_\gamma G_{\alpha\gamma} R_{\gamma\lambda} 
  = \delta_{\alpha\lambda} -\frac1N
  \:.
\end{align}
Since we will have to consider differences between potentials at various contacts, the relation
\begin{align}
  \label{convunit}
  \sum_{\gamma}(R_{\alpha\gamma}-R_{\beta\gamma})\,G_{\gamma\lambda}
  &=\sum_{\gamma}G_{\lambda\gamma}\,(R_{\gamma\alpha}-R_{\gamma\beta})
  \nonumber
  \\
  &=\delta_{\alpha\lambda}-\delta_{\beta\lambda}
\end{align}
will be useful.

\subsection{Four-terminal resistance}

The four-terminal resistance provides information in the situation where current only flows through contacts $\alpha$ and $\beta$, $I_\alpha=-I_\beta=I$, with all other currents vanishing, Eq.~\eqref{eq:Def4TR}.
The voltage at contact $\mu$ is thus conveniently expressed in terms of the resistance matrix as $V_\mu=(R_{\mu\alpha}-R_{\mu\beta})I$ and the voltage difference $V_\mu-V_\nu=\mathcal{R}_{\alpha\beta,\mu\nu}I$ thus involves
\begin{equation}
  \label{4tr-rm}
  \mathcal{R}_{\alpha\beta,\mu\nu} = 
  R_{\mu\alpha}-R_{\mu\beta}-R_{\nu\alpha}+R_{\nu\beta}
  \:.
\end{equation}
A useful relation is obtained by differentiating the relation (\ref{convunit})~\cite{KanLeeDiv88}~:
\begin{equation}
  \label{di4tr}
  \delta\mathcal{R}_{\alpha\beta,\mu\nu}
  =-\sum_{\gamma,\lambda}
  (R_{\mu\gamma}-R_{\nu\gamma})\,\delta G_{\gamma\lambda}
  (R_{\lambda\alpha}-R_{\lambda\beta})
  \:.
\end{equation}


\section{Averaging functions of the conductances}
\label{appendix:proof}

As we have discussed in the body of the text, the quantum contributions to the transport coefficients are naturally expressed for the \textit{conductance} matrix, as linear response theory provides formulae for the conductivity or the conductance.
On the other hand, resistances, which are in general complicated functions of the set of all conductance matrix elements, are more easy to handle in most situations, and usually the relevant quantities in most experiment.
We show in this appendix how one can go from the quantum contributions to the conductance matrix elements to the equivalent contributions to the four-terminal resistances.

Let us consider a general quantity, function of the conductance matrix elements $\mathcal{A}=f(G)$, where $f(G)$ is a short notation for a function of all matrix elements $f(G_{11},G_{12},G_{13},\cdots)$~; the quantity $\mathcal{A}$ may be for example a four-terminal resistance. 
We now show how the quantum contributions to the quantity $\mathcal{A}$ can be related to the quantum contributions to the conductance.
For this purpose it is convenient to analyze the scaling with the number of channels $N_c$, which is a large parameter.~\footnote{
  For a simple geometry, the relevant large parameter is rather the dimensionless conductance $g$ (which is simply $g=\xiloc/L$ for a wire of length $L$, with $\xiloc\propto N_c\ell_e$), however in a complex network of metallic wires, involving several lengths, there is no unique such parameter.
}
Considering the set of conductance matrix elements $\{G_\ab\}$, the disorder average is given by two contributions~: 
a classical term and the weak localisation correction~:
$\smean{G_{\alpha\beta}}=G^\mathrm{class}_{\alpha\beta}+\Delta G_\ab$.
Other quantum contributions of interest are the mesoscopic (sample to sample) fluctuations, defined by $G_\ab=\smean{G_\ab}+\delta G_\ab$ and characterized by the correlation functions $\smean{\delta G_{\alpha\beta}\,\delta G_{\mu\nu}}$.
As we have recalled in the text, the three quantities present the $N_c$ dependences
\begin{align}
  G^\mathrm{class}_{\alpha\beta}                    &= \mathcal{O}(N_c)   \\
  \Delta G_\ab                                      &= \mathcal{O}(N_c^0) \\
  \smean{\delta G_{\alpha\beta}\,\delta G_{\mu\nu}} &= \mathcal{O}(N_c^0)
  \:.
\end{align}
We now deduce two useful properties.

\subsection{Property 1 }
Writing $G_\ab=\smean{G_\ab}+\delta G_\ab$, the fluctuation is smaller than the average, $\delta G_\ab=\mathcal{O}(N_c^0)$, and vanishes on average by definition. As a consequence
\begin{equation}
  \label{eq:Property1}
  \mean{f(G)} = f(\smean{G}) +\mathcal{O}(f\times N_c^{-2})
  \:.
\end{equation}

\subsection{Property 2 }

Using this property, we can split the average into a classical part and the WL correction~:
$\smean{G}=G^\mathrm{class}+\Delta G$.
Hence the average of the physical quantity takes the form
\begin{align}
  &\smean{\mathcal{A}} 
  =f(G^\mathrm{class}+\Delta G) +\mathcal{O}(f\times N_c^{-2})\\
  &= f(G^\mathrm{class}) + \sum_{\mu,\nu}\derivp{ f(G^\mathrm{class}) }{ G_{\mu\nu}^\mathrm{class} }
  \Delta G_{\mu\nu} + \mathcal{O}(f\times N_c^{-2})
  \nonumber
  \:.
\end{align}
We identity the second term as the weak localisation correction to $\mathcal{A}$~:
\begin{equation}
  \label{eq:Property2}
  \Delta\mathcal{A}=
  \sum_{\alpha,\beta}\derivp{\mathcal{A}^\mathrm{class}}{G_\ab^\mathrm{class}}
  \Delta G_\ab
  \:.
\end{equation}

\subsection{Property 3 }

If we consider two quantities $\mathcal{A}$ and $\mathcal{B}$ functions of the transmissions, following the same lines, we deduce the expression for the correlation function
\begin{equation}
  \label{eq:Property3}
  \mean{\delta\mathcal{A}\delta\mathcal{B}}=
  \sum_{\alpha,\beta,\mu,\nu}
  \derivp{\mathcal{A}^\mathrm{class}}{G_\ab^\mathrm{class}}
  \derivp{\mathcal{B}^\mathrm{class}}{G_\mn^\mathrm{class}}
  \mean{\delta G_\ab\delta G_\mn}
  \:.
\end{equation}

\subsection{Applications}
\label{app:Dem4TRF}

\noindent$\bullet$ {\bf WL correction to the four-terminal resistances~:}
It is now straightforward to get the weak localisation correction to the 
four-terminal resistance. We deduce from the property \eqref{eq:Property1}\footnote{
  Whereas the classical resistances and 
  conductances satisfy the relation
  $\sum_\beta R_\ab^\mathrm{class} G_{\beta\lambda}^\mathrm{class} 
   = \delta_{\alpha\lambda} -1/N$,
  the average resistance and conductance do not, up to correction of order
  $N_c^{-2}$ due to the correlations between transmissions.
  However the relation is obviously satisfied for a given 
  realization of disorder~:
  $\sum_\beta R_\ab G_{\beta\lambda} = \delta_{\alpha\lambda} -1/N$.
}
$
\sum_{\gamma}
( \smean{R_{\alpha\gamma}} - \smean{R_{\beta\gamma}} )\,
\smean{G_{\gamma\lambda}}
=\delta_{\alpha\lambda}-\delta_{\beta\lambda} + \mathcal{O}(N_c^{-2})
$.
Then expanding the conductances as 
$\smean{G_{\gamma\lambda}}=G_{\gamma\lambda}^\mathrm{class}+\Delta G_{\gamma\lambda}$, 
and using (\ref{di4tr}) and the property \eqref{eq:Property2}, one get
\begin{align}
  &\Delta\mathcal{R}_{\alpha\beta,\mu\nu}  
  \\ \nonumber
  &= - \sum_{\gamma,\lambda}
  (R_{\mu\gamma}^\mathrm{class}-R_{\nu\gamma}^\mathrm{class})\,\Delta G_{\gamma\lambda}
  (R_{\lambda\alpha}^\mathrm{class}-R_{\lambda\beta}^\mathrm{class})
  \:.
\end{align}
Using (\ref{eq:TexMonPRL04}) and once again the equation (\ref{di4tr}) we finally get \eqref{eq:WL4TR}.
Note that this result could have been guessed from the heuristic argument presented in the introduction of Ref.~\cite{TexMon04}.

\vspace{0.25cm}

\noindent$\bullet$ {\bf Correlations of four-terminal resistances~:}
The expressions for the correlation functions (\ref{eq:4TRcorrel1},\ref{eq:4TRcorrel2},\ref{eq:4TRcorrel3},\ref{eq:4TRcorrel4}) are demonstrated by making use of the relation \eqref{di4tr} with the property \eqref{eq:Property3} (with the remark closing paragraph~\ref{subsec:4.1}).


\section{Solution of the diffusion equation in the star graph}
\label{appendix:Star}

The solution of the diffusion equation $\big[\gamma-\partial_x^2\big]P_c(x,x')=\delta(x-x')$
in the star graph with $m_\alpha$ infinite wires is useful (Fig.~\ref{fig:star-graph}).
Details are given in Appendix D of Ref.~\cite{TexDelMon09}.
The value of the Cooperon at the vertex $\alpha$ is inversely proportional to the coordination number of the vertex~: $P_c(\alpha,\alpha)=1/(m_\alpha\sqrt{\gamma})$.
When the two arguments belong to  wires $x\in(i)$ and $x'\in(j)$, we get
\begin{align}
  P_c(x,x') =& \frac{1}{m_\alpha\sqrt{\gamma}}\,\EXP{-\sqrt{\gamma}(x+x')}
  \nonumber\\
  &+\delta_{i,j}\frac{1}{\sqrt{\gamma}}\sinh(\sqrt{\gamma}x_<)\,\EXP{-\sqrt{\gamma}x_>}
  \:,
\end{align}
where $x_<=\mathrm{min}(x,x')$, $x_>=\mathrm{max}(x,x')$ and the distance is measured from the vertex.
At large distance of the vertex, we recover the result for the infinite wire
$P_c(x,x')=[1/(2\sqrt{\gamma})]\exp\big[-\sqrt{\gamma}|x-x'|\big]$.

\begin{figure}[!ht]
\centering
\includegraphics[width=0.25\textwidth]{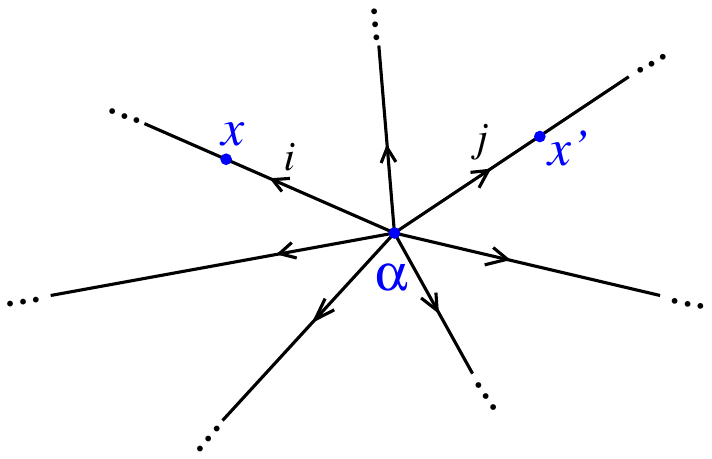}
\caption{\it A star graph with coordination number $m_\alpha=7$.}
\label{fig:star-graph}
\end{figure}

\end{appendix}





\end{document}